\documentclass[twocolumn]{aastex63}
\pdfoutput=1
\usepackage{physics}
\usepackage{graphicx}
\usepackage{enumerate}
\usepackage{amssymb, amsmath}
\usepackage{natbib}
\usepackage{color}
\usepackage{hyperref}

\newcommand{\Gaia}{{Gaia}}

\newcommand{\Msun}{\mbox{$M_{\sun}$}}
\newcommand{\Mearth}{\mbox{$M_{\oplus}$}}

\newcommand{\Rsun}{\mbox{$R_{\sun}$}}
\newcommand{\Mjup}{\mbox{$M_{\rm Jup}$}}

\defcitealias{Artymowicz-Lubow1994}{AL94}
\defcitealias{Ueda2017}{U17}

\usepackage{xspace}
\newcommand{\codename}{{\tt orvara}\xspace}
\newcommand{\gaia}{{Gaia}\xspace}
\newcommand{\hipparcos}{{Hipparcos}\xspace}
\newcommand{\MB}{$0.5425\pm 0.0042\,\Msun$\xspace}
\newcommand{\msini}{$4.266_{-0.087}^{+0.11}\,\Mjup$\xspace}
\newcommand{\MAnounit}{$0.870_{-0.026}^{+0.035}$\xspace}
\newcommand{\MBnounit}{$0.5425\pm 0.0042$\xspace}
\newcommand{\msininounit}{$4.266_{-0.087}^{+0.11}$\xspace}
\newcommand{\MAnoerr}{$0.870\,\Msun$\xspace}

\newcommand{\anow}{$23.7\pm 0.3$\,AU\xspace}
\newcommand{\anownounit}{$23.7\pm 0.3$\xspace}
\newcommand{\anownoerror}{$23.7$\,AU\xspace}
\newcommand{\aprimordial}{$14.8$\,AU\xspace}
\newcommand{\aAGB}{$14.9$\,AU\xspace}

\newcommand{\e}{$0.429\pm 0.017$\xspace}
\newcommand{\enoerr}{$0.429$\xspace}
\newcommand{\inounit}{$126.44_{-0.49}^{+0.47}$\xspace}
\newcommand{\Omeganounit}{$234.2\pm 1.0$\xspace}

\newcommand{\aplanet}{$0.1177_{-0.0012}^{+0.0015}$\xspace}
\newcommand{\eplanet}{$0.0478\pm 0.0024$\xspace}

\newcommand{\Rrim}{\mbox{$R_{\rm rim}$}}
\newcommand{\Msunperyr}{\mbox{$\Msun$\,yr$^{-1}$}}
\newcommand{\mustar}{\mbox{$\mu_{\star}$}}
\newcommand{\microm}{\mbox{$\rm \mu m$}}

\newcommand\numberthis{\addtocounter{equation}{1}\tag{\theequation}}
\usepackage{afterpage}
\usepackage[english]{babel}

\begin{document}

\title{The Gliese 86 Binary System: A Warm Jupiter Formed in a Disk Truncated at $\approx$2 AU}

\author[0000-0003-4594-4331]{Yunlin Zeng}
\affiliation{School of Physics, Georgia Institute of Technology, Atlanta, GA 30332, USA}

\author[0000-0003-2630-8073]{Timothy D.~Brandt}
\affiliation{Department of Physics, University of California, Santa Barbara, Santa Barbara, CA 93106, USA}

\author[0000-0001-8308-0808]{Gongjie Li}
\affiliation{Center for Relativistic Astrophysics, School of Physics, Georgia Institute of Technology, Atlanta, GA 30332, USA}

\author[0000-0001-9823-1445]{Trent J.~Dupuy}
\affiliation{Institute for Astronomy, University of Edinburgh, Royal Observatory, Blackford Hill, Edinburgh, EH9 3HJ, UK}

\author[0000-0002-6845-9702]{Yiting Li}
\affiliation{Department of Physics, University of California, Santa Barbara, Santa Barbara, CA 93106, USA}

\author[0000-0003-0168-3010]{G.~Mirek Brandt}
\altaffiliation{NSF Graduate Research Fellow}
\affiliation{Department of Physics, University of California, Santa Barbara, Santa Barbara, CA 93106, USA}

\author[0000-0003-1748-602X]{Jay Farihi}
\affiliation{Department of Physics and Astronomy, University College London, London WC1E 6BT, UK}

% \author[0000-0001-5415-9189]{Gail Schaefer}
% \affiliation{The CHARA Array of Georgia State University, Mount Wilson Observatory, Mount Wilson, CA 91023, USA}

\author[0000-0002-1160-7970]{Jonathan Horner}
\affiliation{Centre for Astrophysics, University of Southern Queensland, Toowoomba, QLD 4350, Australia}

\author[0000-0001-9957-9304]{Robert A.~Wittenmyer}
\affiliation{Centre for Astrophysics, University of Southern Queensland, Toowoomba, QLD 4350, Australia}

\author[0000-0003-1305-3761]{R.~Paul. Butler}
\affiliation{Earth and Planets Laboratory, Carnegie Institution for Science, Washington, DC 20015, USA}

\author[0000-0002-7595-0970]{Christopher G.~Tinney}
% \affiliation{Exoplanetary Science at University of New South Wales, School of Physics, University of New South Wales, Sydney, NSW 2052, Australia}
\affiliation{Exoplanetary Science at UNSW, School of Physics, UNSW Sydney, Sydney, NSW 2052, Australia}
\affiliation{Australian Centre for Astrobiology, UNSW Sydney, Sydney, NSW 2052, Australia}

\author[0000-0003-0035-8769]{Bradley D.~Carter}
\affiliation{Centre for Astrophysics, University of Southern Queensland, Toowoomba, QLD 4350, Australia}

\author[0000-0001-7294-5386]{Duncan J.~Wright}
\affiliation{Centre for Astrophysics, University of Southern Queensland, Toowoomba, QLD 4350, Australia}

\author[0000-0003-0433-3665]{Hugh R.~A.~Jones}
\affiliation{Centre for Astrophysics Research, University of Hertfordshire, Hatfield AL10 9AB, UK}

\author[0000-0003-2839-8527]{Simon J.~O’Toole}
\affiliation{Australian Astronomical Optics, Macquarie University, North Ryde, NSW 1670, Australia}

\begin{abstract}  
Gliese 86 is a nearby K dwarf hosting a giant planet on a $\approx$16-day orbit and an outer white dwarf companion on a $\approx$century-long orbit. In this study we combine radial velocity data (including new measurements spanning more than a decade) with high angular resolution imaging and absolute astrometry from \hipparcos and \gaia to measure the current orbits and masses of both companions.  We then simulate the evolution of the Gl~86 system to constrain its primordial orbit when both stars were on the main sequence; the closest approach between the two stars was then about $9\,$AU. Such a close separation limited the size of the protoplanetary disk of Gl~86~A and dynamically hindered the formation of the giant planet around it. Our measurements of Gl~86~B and Gl~86~Ab's orbits reveal Gl~86 as a system in which giant planet formation took place in a disk truncated at $\approx$2\,AU. Such a disk would be just big enough to harbor the dust mass and total mass needed to assemble Gl~86~Ab's core and envelope, assuming a high disk accretion rate and a low viscosity. Inefficient accretion of the disk onto Gl~86~Ab, however, would require a disk massive enough to approach the Toomre stability limit at its outer truncation radius.  The orbital architecture of the Gl~86 system shows that giant planets can form even in severely truncated disks and provides an important benchmark for planet formation theory. 
\end{abstract}

\section{Introduction}

Thousands of exoplanets are now known in a huge variety of systems, and in an enormous range of dynamical configurations \citep{Luger2017, Shallue2018, Lam2020}.  These include hot Jupiters \citep{Butler1997, Henry2000, Tinney2001}, outer Jovian planets \citep{Jones2010, Wittenmyer2014, Venner2021}, smaller planets of all sizes and orbital distances \citep{Barclay2013, Jenkins2015, Smith2018}, planets around binaries \citep[P-type systems;][]{Welsh2012, Orosz2019, Kostov2021}, and planets around individual stars within binaries \citep[S-type systems;][]{Teske2016, Hatzes2003, Ramm2016}.  This diversity suggests that planet formation is a robust, if not a universal, process accompanying star formation.

Planet formation in binaries is an especially important testbed for the planet formation process.  The existence of the binary provides natural constraints on the properties of circumstellar and circumbinary disks, and therefore on the material available for planet formation.  Both P- and S-type planets must form within a disk, but one that is dynamically interacting with the binary in an environment very different from the canonical Solar nebula.

\cite{Su2021} conducted a statistical study of the S-type planetary systems detected from radial velocity (RV) surveys to generalize the characteristics of these systems. Table~1 of that paper summarizes the properties of $80$ planet-hosting binaries; ten of them (HD~42936, HD~87646, HD~59686, HD~7449, $\gamma$~Cep, HD~4113, HD~41004, 30~Ari, Gl~86, and HD~196885) have separations smaller than $30$\,AU. \cite{Jang-Condell2015} argued that the frequent appearance of the planets in close binaries indicates the formation process is robust. However, that the binaries are close to each other limits the amount of materials in the circumstellar disk and significantly reduces the chance of forming planetary embryos. To better understand the planet formation under such conditions, we focus on Gl~86 and investigate its orbital parameters and possible planet formation scenarios in this paper. 

Gl~86 is the second-closest planetary system containing a warm or hot Jupiter, after Gl~876, with a distance of $10.761\pm0.005$\,pc \citep{GaiaEDR3-ast-sol} and an age of $\approx$10\,Gyr \citep{Fuhrmann2014}. \citet{Queloz2000} discovered this system using the CORALIE echelle spectrograph and found an RV signal corresponding to an $m\sin{i} \approx 4\,\Mjup$ planet with a 15.8-day orbital period as well as a distant and massive companion causing a long-term RV drift. \citet{Els2001} used the ADONIS adaptive optics system on the ESO 3.6-m Telescope at La Silla to observe Gl~86 and identified a wide companion that they inferred to be a brown dwarf causing the RV drift. \citet{Mugrauer2005} performed additional high-contrast observations and found that this wide companion is, instead, a cool white dwarf, and they ruled out any additional stellar companions between $0\farcs1$ and $2\farcs1$, or $1$--$23$\,AU. \citet{Lagrange2006} used VLT/NACO to obtain photometric and astrometric measurements and confirmed that the companion is a white dwarf rather than a brown dwarf and inferred its mass to be $0.48\,\Msun \leq m \leq 0.62\,\Msun$ based on the amplitude of the RV trend observed by \citet{Queloz2000}.
\citet{Brandt2019} used all of the above measurements, as well as additional relative astrometry from \citet{Farihi2013} and the proper motion anomaly between \hipparcos and \gaia DR2 \citep{Brandt2018}, to determine the mass and orbital parameters of the white dwarf.

The Gl~86 system, with a white dwarf on a $\approx$20-AU orbit and a close-in, gas-giant exoplanet, challenges planet formation models. 
From a theoretical perspective, such close binary systems are expected to be hostile to the formation of giant planets due to disk truncation \citep[e.g.,][]{Artymowicz-Lubow1994} and destructive planetesimal collisions \citep[e.g.,][]{Paardekooper2010, Rafikov2015}, and this is largely borne out by observations \citep[e.g.,][]{Wang2014, Kraus2016}.
The Gl~86 system presents a further problem because when both stars were on the main sequence, the separation between them was even smaller and the stability and the feasibility of forming the inner planet becomes even more questionable. Both \citet{Lagrange2006} and \citet{Farihi2013} doubted the orbital stability of the inner planet since the semi-major axis of the primordial binary system was too small. In order to work out a theory regarding the formation of the warm Jupiter in the Gl~86 system, it is necessary to better constrain Gl~86's current and primordial orbital parameters, and the stellar masses.

In this paper, we perform a new fit to the masses and orbits of the Gl~86 system using absolute astrometry from the latest Gaia Data Release \citep[EDR3,][]{Gaia-EDR3-summary, GaiaEDR3-ast-sol}, together with RV and relative astrometry data from the literature. We discuss the resulting orbital elements of the Gl~86 system in Section~\ref{sec: current orbit}. In Section~\ref{sec: primordial orbit}, we simulate the binary's evolution based on an $N$-body integrator program in order to constrain its primordial orbit. In Section~\ref{sec:formation}, we discuss some implications that the primordial orbit has on the formation of the planet surrounding Gl~86~A and the challenges the planet faced by the time it was formed. (Because Gl~86~B starts with a higher mass and ends with a lower mass due to mass loss, we consistently call Gl~86~A the host and Gl~86~B the companion stars, instead of primary and secondary, to avoid confusion.) Finally, Section \ref{sec:conclusions} summarizes our results.

\section{The current orbit of Gl~86} \label{sec: current orbit}

We use the open-source python package \codename \citep{Brandt2021} to fit for the current masses and orbits in the Gl~86 system.  The program can fit any combination of RVs, relative, and absolute astrometry from \hipparcos and \Gaia.  We use all of these types of data to constrain Gl~86.  In this section, we describe the input data and our resulting fit.

\subsection{Data}

We use the absolute astrometry from \hipparcos \citep{Hipparcos_1997,vanLeeuwen_2007} and \gaia's latest data release \citep{Gaia-EDR3-summary} as cross-calibrated by \cite{Brandt2021_HGCA}. The \hipparcos-\gaia Catalog of Accelerations \citep[HGCA,][]{Brandt2018,Brandt2021_HGCA} provides three proper motions on the \gaia EDR3 reference frame; differences between them indicate astrometric acceleration.  For Gl~86, the two most precise proper motions are the one computed from the position difference between \hipparcos and \gaia and the \gaia EDR3 proper motion.  These two measurements are inconsistent with constant proper motion at nearly $300\sigma$ significance.

We use relative astrometry from multiple literature sources.  Table \ref{tab: relAst} lists our adopted relative astrometry for Gl~86~B, where $\rho$ is the separation between the two stars and PA is the position angle (east of north). The last data point in Table \ref{tab: relAst} was imaged on 2016 November 10 using the Space Telescope Imaging Spectrograph (STIS) as part of program GO-14076 (PI G\"ansicke). The imaging was performed using the narrow-band filter F28X50OII (central wavelength 3738\,\AA, full width at half maximum 57\,\AA) with a series of four two-second exposures in a standard dither pattern. This filter has no red leak, and thus the bright host star remains unsaturated and in the linear response regime. The white dwarf companion is also detected in all four frames, at approximate signal-to-noise ratios between 13 and 19. This set of images was used to robustly measure the separation of the binary, where the companion star was found at offset $2.\!\!^{\prime\prime}6220 \pm 0.\!\!^{\prime\prime}0040$ with position angle $82.\!\!^\circ185\pm 0.\!\!^\circ098$ under the J2000 frame.

We note that the relative position measurement of the two stars in Gaia EDR3 is less straightforward than the other relative astrometry measurements. It is the difference between the 2016.0 positions of five-parameter astrometric solutions to each star in the binary.  The formal uncertainties are tiny, but are subject to possible systematics from the proximity of the two stars and from their straddling of the $G = 13$\,mag boundary where the window function changes \citep{Gaia-EDR3-summary,Cantat-Gaudin+Brandt_2021}.  We treat the measurement as instantaneous and, somewhat arbitrarily, adopt uncertainties similar to the HST uncertainties.  This avoids having Gaia solely drive the result and mitigates the possible impact of systematics.  

The RV data come from the UCLES \'echelle spectrograph \citep{Diego1990} on the Anglo-Australian Telescope. Those results spanning 1998 to 2005 were published in \citet{Butler2006}.  We also include a further 34 previously unpublished UCLES RVs spanning 2006 to 2015 (Table \ref{tab: RV}), for a total time baseline of 17.8 years.  The RV data are processed through the same pipeline, but they have some discrepancy with \citet{Butler2006} due to minor pipeline tweaks and the fact that they have the mean stellar RV subtracted.  Thanks to Gl~86~A's very large acceleration, the mean RV has changed appreciably with an additional nine years of data. 
%since they are computed relative to the stellar template and the mean changes for every observation.

\begin{deluxetable*}{cccc}
\tablewidth{0pt} %[htbp]
%\centering
 \caption{Direct imaging astrometry of Gl~86 \label{tab: relAst}
}
% \begin{tabular}{c c c c} 
% \hline
% \hline
\tablehead{
  Date (Jyear) & 
  $\rho$ (arcsec) & 
  PA (degrees) & 
  Reference
  } 
% \hline
\startdata
  2000.82 & $1.73 \pm 0.03$ & $119 \pm 1$ & \citet{Els2001} \\ 
  2003.87 & $1.906 \pm 0.015$ & $107.5 \pm 0.5$ & \citet{Lagrange2006} \\
  2004.73 & $1.941 \pm 0.014$ & $105.3 \pm 0.6$ & \citet{Lagrange2006} \\
  2005.03 & $1.93 \pm 0.02$ & $104.0 \pm 0.4$ & \citet{Mugrauer2005} \\
  2005.57 & $1.969 \pm 0.011$ & $102.7 \pm 0.4$ & \citet{Lagrange2006} \\
  2012.2468 & $2.351 \pm 0.002$ & $88.96 \pm 0.04$ & \citet{Farihi2013} \\
  2016.0\tablenotemark{a}  & $2.5725 \pm 0.0020$ & $83.36 \pm 0.10$ & \citet{GaiaEDR3-ast-sol} \\ %(eDR3)
  2016.8606 & $2.6220 \pm 0.0040$ & $82.19 \pm 0.10$ & STIS, this work 
  \enddata
\tablenotetext{a}{Measurement is not instantaneous}
\end{deluxetable*}

%\LongTables
\begin{deluxetable}{ccc}
\tablewidth{0pt} %[htbp]
 \caption{RV data of Gl~86. All RV are available electronically. % \textcolor{blue}{(To the Journal: We'd like to include the rest of the data in the supplementary online paper (commented out in the Latex source code).)} 
 \label{tab: RV}
}

\tablehead{
  Date & 
  RV (m/s) & 
  $\sigma_{\rm RV} \rm (m/s)$
  } 
\startdata
2450831.03498  &   958.65   &  1.66 \\
2451211.96513  &  1227.68   &  2.17 \\
2451213.98147  &  1282.40   &  2.28 \\
2451214.92978  &  1227.60   &  1.94 \\
2451235.93120  &   601.25   &  1.92 \\
... & ... & ...
  \enddata
\end{deluxetable}

\subsection{Orbital Fit}

We use \codename to fit a superposition of Keplerian orbits to the Gl~86~A astrometry and RVs and relative astrometry.  We use log-uniform priors for semimajor axis and companion mass, a geometric prior on inclination, and uniform priors on the remaining orbital parameters.  We adopt the \gaia EDR3 parallax as our parallax prior; \codename analytically marginalizes parallax out of the likelihood.  We use a log-uniform prior on RV jitter.  

We adopt an informative prior on the mass of Gl~86~A. \citet{Brandt2019} obtained $M_{\rm A} = 1.39_{-0.23}^{+0.24}\, \Msun$ by using the cross-calibrated \hipparcos and \gaia DR2 astrometry in a fit to Gl~86~B. Their prior was log-flat, but stellar evolution allows a much narrower prior.  
\citet{Fuhrmann2014} modelled Gl~86~A's atmosphere using high-resolution spectroscopy and concluded $M_{\rm A} = 0.83 \pm 0.05\, \Msun$. We adopt this as our prior on Gl~86~A's mass.

The log likelihood function consists of three parts: the chi-squared values of RV (including a penalty term for RV jitter), relative astrometry, and absolute astrometry. To maximize the likelihood is equivalent to minimizing the negative log likelihood,
\begin{equation}
    -2\ln \mathcal{L} = \chi^2 = \chi^2_{\rm RV} + \chi^2_{\rm rel\,ast} + \chi^2_{\rm abs\,ast}.
\end{equation}
In addition, the RV zero point, parallax, and the proper motion of the system's barycenter are marginalized out as nuisance parameters. We refer readers to the \codename paper for more details on the formulas and techniques used.

We use Markov Chain Monte Carlo (MCMC) to explore the posterior probability distribution with the {\tt emcee} \citep{Foreman-Mackey_2013} and {\tt ptemcee} \citep{Vousden+Farr+Mandel_2016} packages.  
Parallel tempering MCMC uses walkers at many temperatures, each multiplying an extra factor of $1/\sqrt{T}$ to the exponent of the likelihood. A larger temperature means a more flattened out posterior probability distribution and enables hotter temperature walkers to explore more parameter space. Temperature swaps can happen periodically while preserving detailed balance. Parallel-tempering helps to explore multimodal posteriors and avoid getting stuck at some local minimum. 
We use $100$ walkers, $30$ temperatures, and $2\times10^5$ steps; we keep every 50th step and use the coldest chain for inference.  We discard the first $1.25\times10^5$ steps as burn-in.  

\subsection{Results} \label{subsec: current orbit results}

Our first step with our resulting chains is to test whether they include formally well-fitting orbits. A satisfactory fit will have each data point contribute $\approx$1 to the total $\chi^2$.  Unfortunately,
our best-fit $\chi^2$ of relative separation is $59.4$, and that of position angle is $50.5$; both are much too large for 8 data points. Our high best-fit $\chi^2$ values show that either we have underestimated uncertainties, or there is an additional component in the system.  Any additional component massive enough to affect the astrometry would have to orbit Gl~86~B to avoid detection in the precision RVs and direct imaging of Gl~86~A.  Bringing the relative astrometry into agreement requires a perturbation of $\sim$10\,mas, which could be caused by a $\sim$20\,$M_{\rm Jup}$ companion on a $\sim$2\,AU orbit.  However, such companions are rare \citep{Marcy+Butler_2000,Halbwachs+Arenou+Mayor+etal_2000}, and we have just two relative astrometry measurements at mas precision.  This is insufficient to constrain the mass and orbit of a hypothetical substellar companion to Gl~86~B. We provisionally attribute the high $\chi^2$ values to a combination of underestimated uncertainties and systematics in the data.  

For our final orbit analysis, we inflate the uncertainties in the absolute astrometry by a factor of $2$, add $10$\,mas to our relative separation uncertainties and $0.05$\,degrees to our position angle uncertainties, both in quadrature, in addition to the error inflation used by \cite{Brandt2019}. This brings the $\chi^2$ values to an acceptable level ($\chi^2_{\rm relsep} = 11.8$ and $\chi^2_{\rm PA} = 11.3$), and has only a minor impact on our derived parameters.
The mass of Gl~86~B changes by just 0.5\%, while the best-fit semimajor axis decreases from $\approx$25\,AU to $\approx$24\,AU and the best-fit eccentricity increases from 0.38 to 0.43.  Table~\ref{table: Gl~86 orbital params} lists the full set of orbital parameters.

\begin{deluxetable}{cc}
\tablewidth{0pt}
\caption{MCMC results. $\Omega$ is the longitude of ascending node. \label{table: Gl~86 orbital params}}
\tablehead{
Parameter & 
Value
}
\startdata
 host star \\
 \hline
  $M_{\rm A}\,(\Msun)$ & \MAnounit \\ 
 \hline
 white dwarf companion \\
 \hline
  $M_{\rm B}\,(\Msun)$ & \MBnounit \\
  $a_{\rm B}$\,(AU) & \anownounit \\
  $e_{\rm B}$ & \e \\
  $i_{\rm B}\,(^{\circ})$ & \inounit \\
%   $\omega$ & $1$ \\
  $\Omega_{\rm B}\,(^{\circ})$ & \Omeganounit \\
 \hline
 inner planet \\
 \hline
  $m_{\rm b}\sin{i_{\rm b}} \,(\Mjup)$ & \msininounit \\
  $a_{\rm b}$\,(AU) & \aplanet \\
  $e_{\rm b}$ & \eplanet 
\enddata
\end{deluxetable}

\begin{figure*} %[p!]
    \centering
    \includegraphics[width=\textwidth]{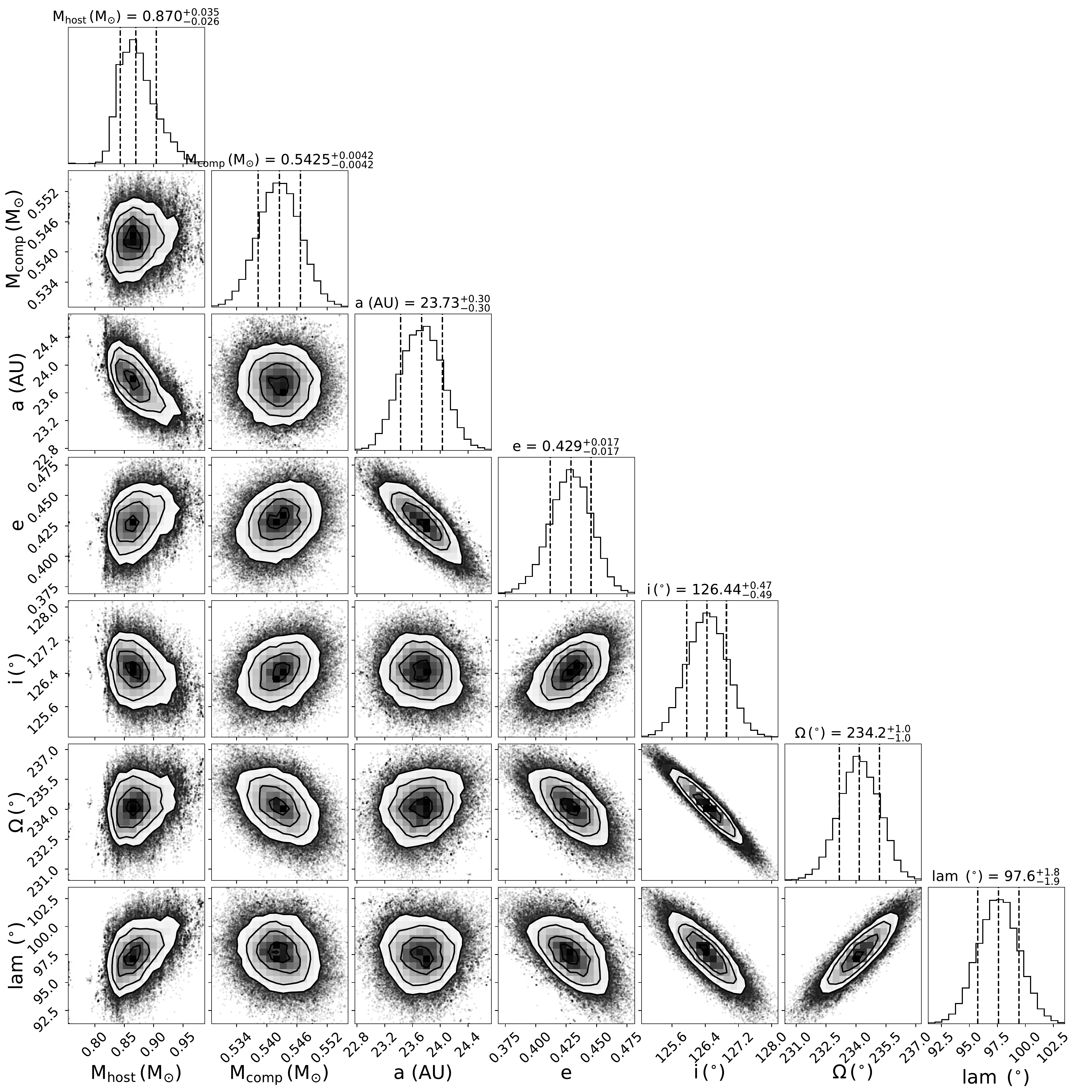}
    \caption{The corner plot of parameters of the white dwarf Gl~86~B. Along the diagonal are the marginalized distribution of each parameter. Others are 2-D joint posterior distribution of each of two parameters. Most parameters are well-constrained, though some are strongly covariant.}
    \label{fig: corner_B}
\end{figure*}

\begin{figure*} %[p!]
    \centering
    \includegraphics[width=\textwidth]{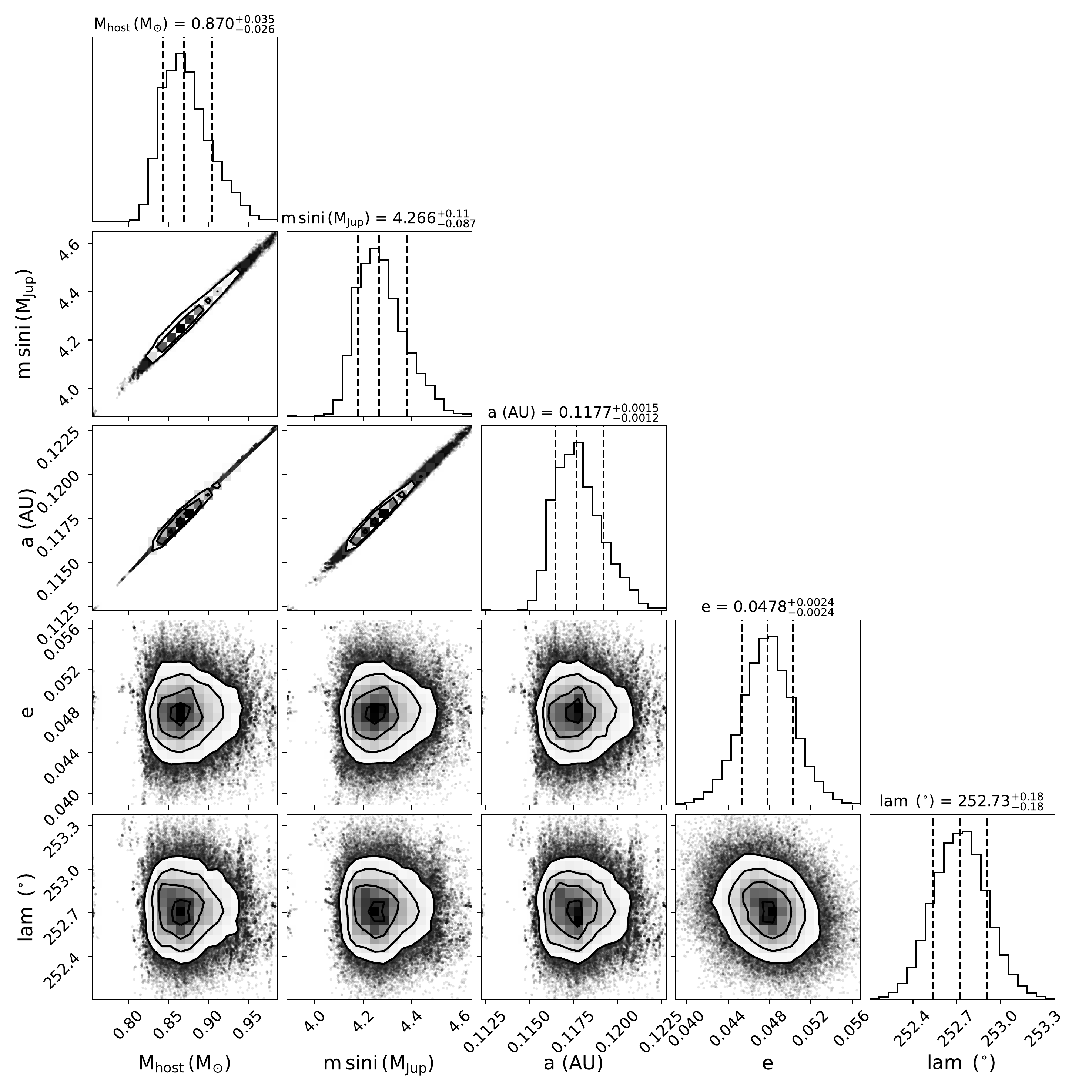}
    \caption{Same as Figure \ref{fig: corner_B} but for the parameters of the planet Gl~86~Ab. The planet's short orbital period prevents a constraint on either inclination or position angle; we plot only $m \sin i$.  The near-total correlation between mass and semimajor axis results from Kepler's Third Law.}
    \label{fig: corner_Ab}
\end{figure*}

The MCMC results are summarized in Figures~\ref{fig: corner_B} and \ref{fig: corner_Ab}. The parameters for the white dwarf companion Gl~86~B are well constrained, but some parameters are strongly correlated. For example, its semi-major axis is anti-correlated with the eccentricity, and its inclination is also anti-correlated with its longitude of ascending node.
The inner planet, Gl~86~Ab, has an orbital period much shorter than either the \hipparcos or \gaia mission baseline.  This, combined with the planet's low $m\sin{i}$, means that we have almost no constraint on its orbital inclination or orientation.  Even epoch astrometry from \gaia DR4 might not detect the $\lesssim$100\,$\mu$as orbit of Gl~86~A about its barycenter with Gl~86~Ab.

\begin{figure}
    \centering
    \includegraphics[width=\linewidth]{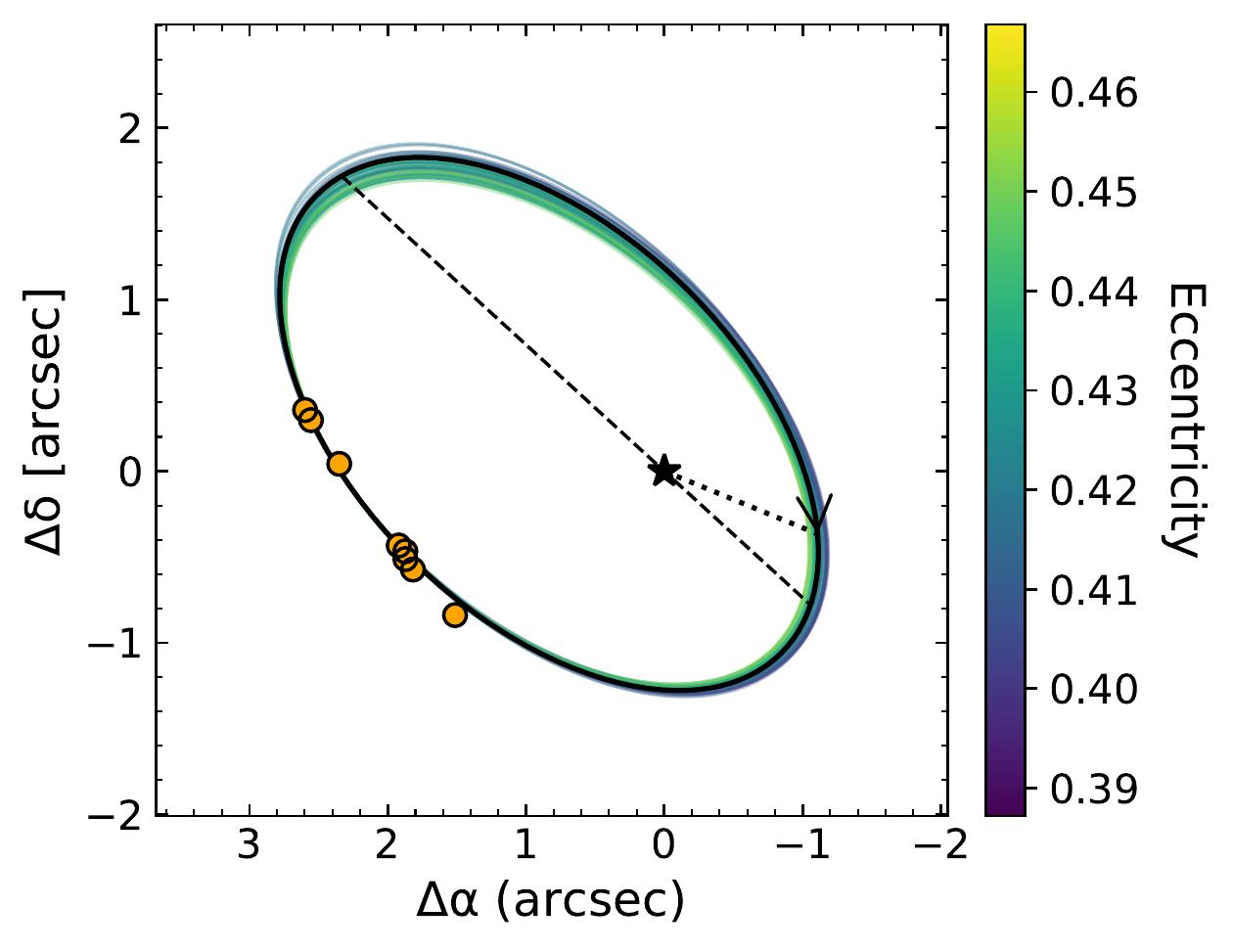}
    \caption{Relative astrometric orbit of Gl~86~AB. The solid black curve indicates the most probable orbit. The colorful curves show 50 randomly selected orbits from the posterior probability distribution, with color denoting eccentricity. The star symbol at the origin represents the host star, the dotted line connects the host star to the periapsis, and the dashed line is the line of nodes. The orange circles are the relative astrometry data points.}
    \label{fig: astrometric_B}
\end{figure}

\begin{figure*}
    \centering
    \includegraphics[width=.49\textwidth]{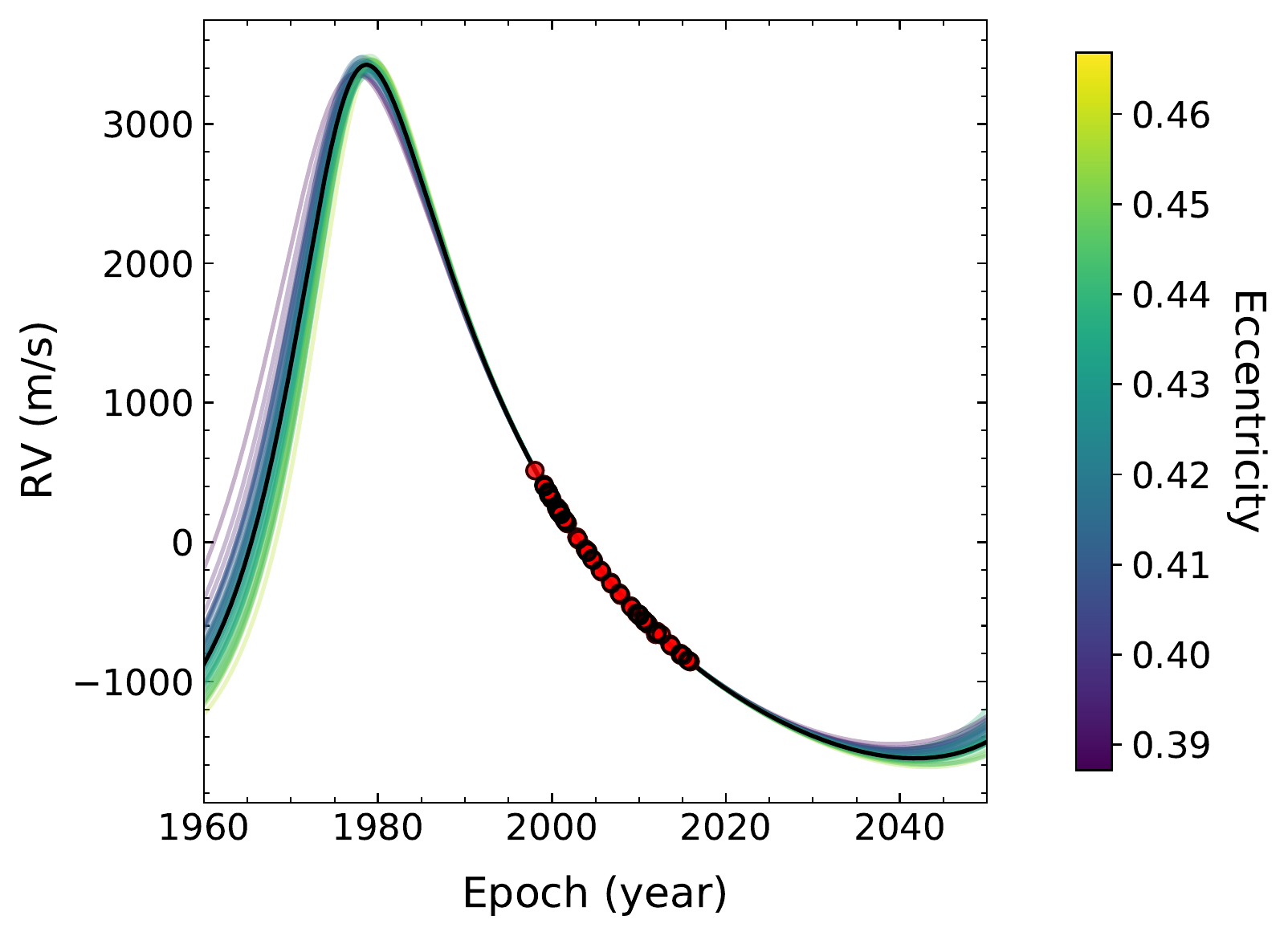}
    \includegraphics[width=.49\textwidth]{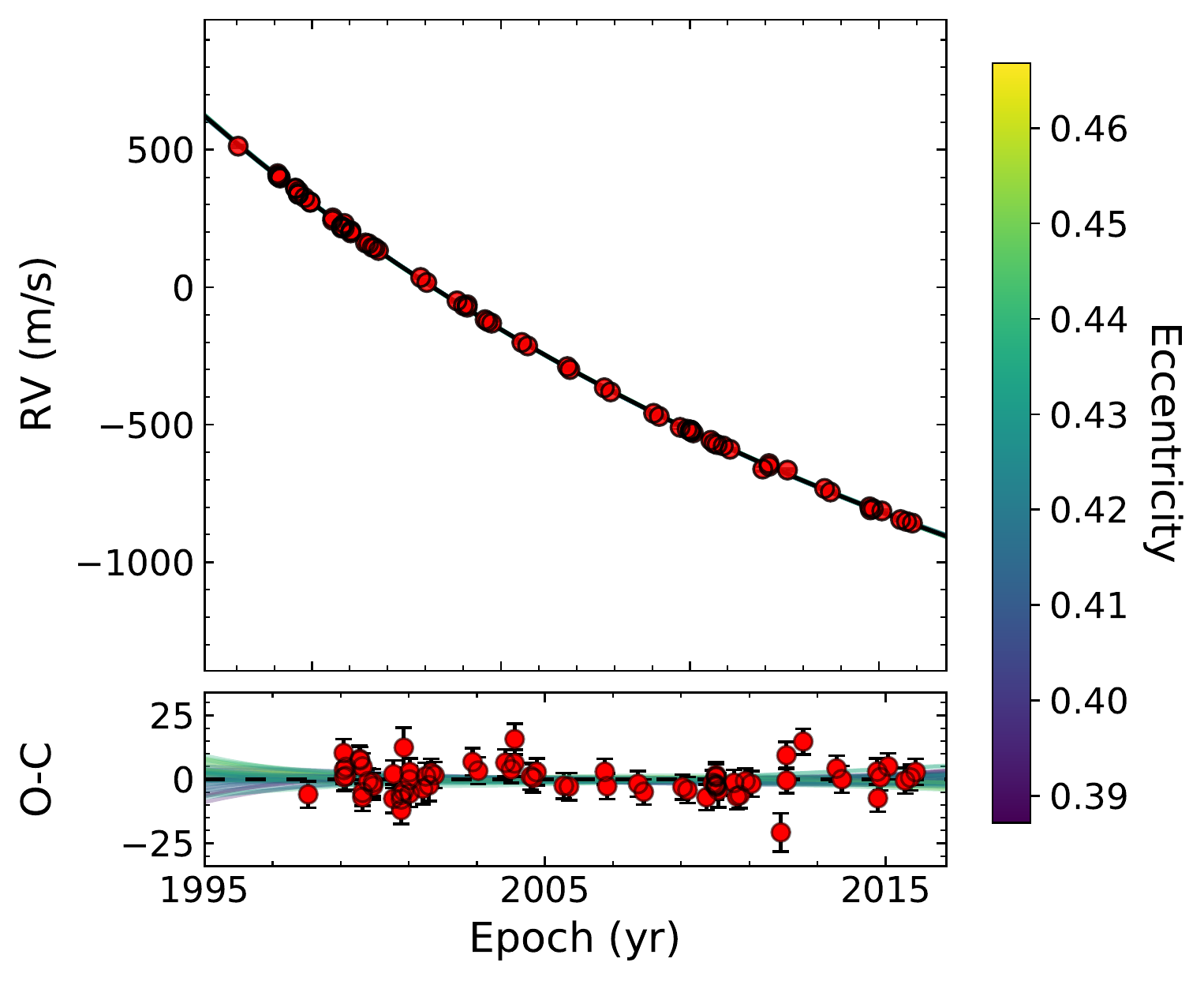}

    \includegraphics[width=.49\textwidth]{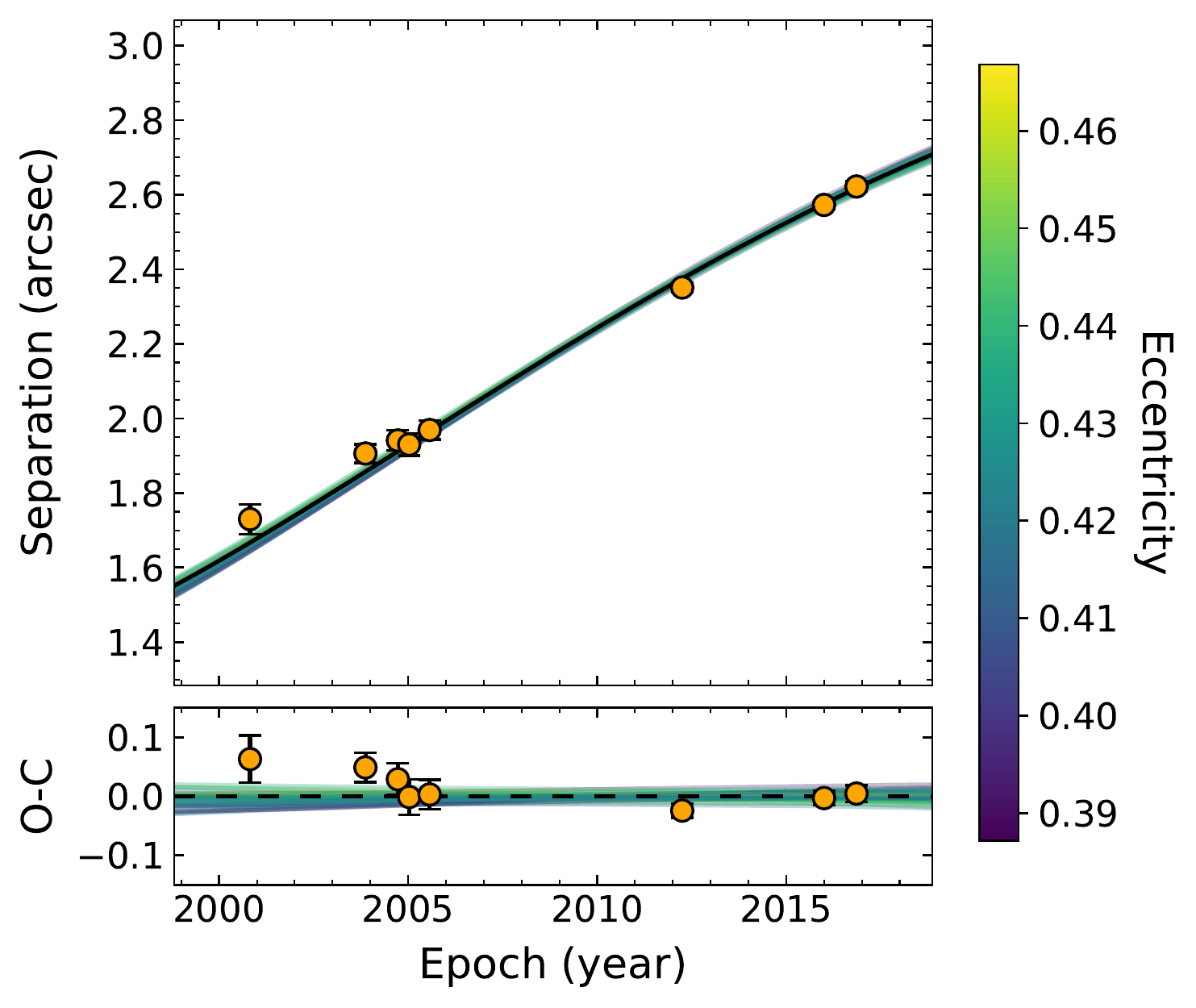}
    \includegraphics[width=.49\textwidth]{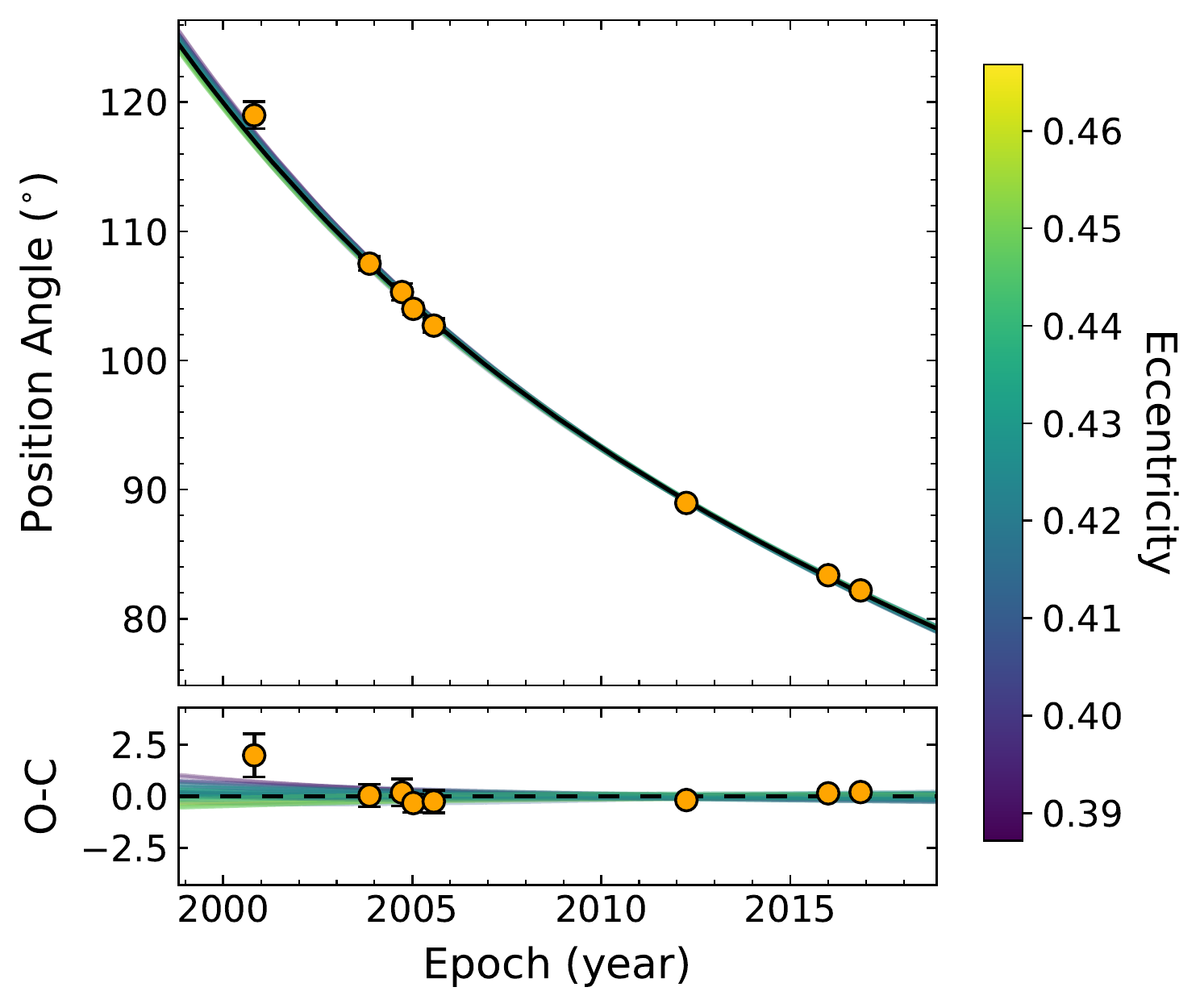}

    \includegraphics[width=.49\textwidth]{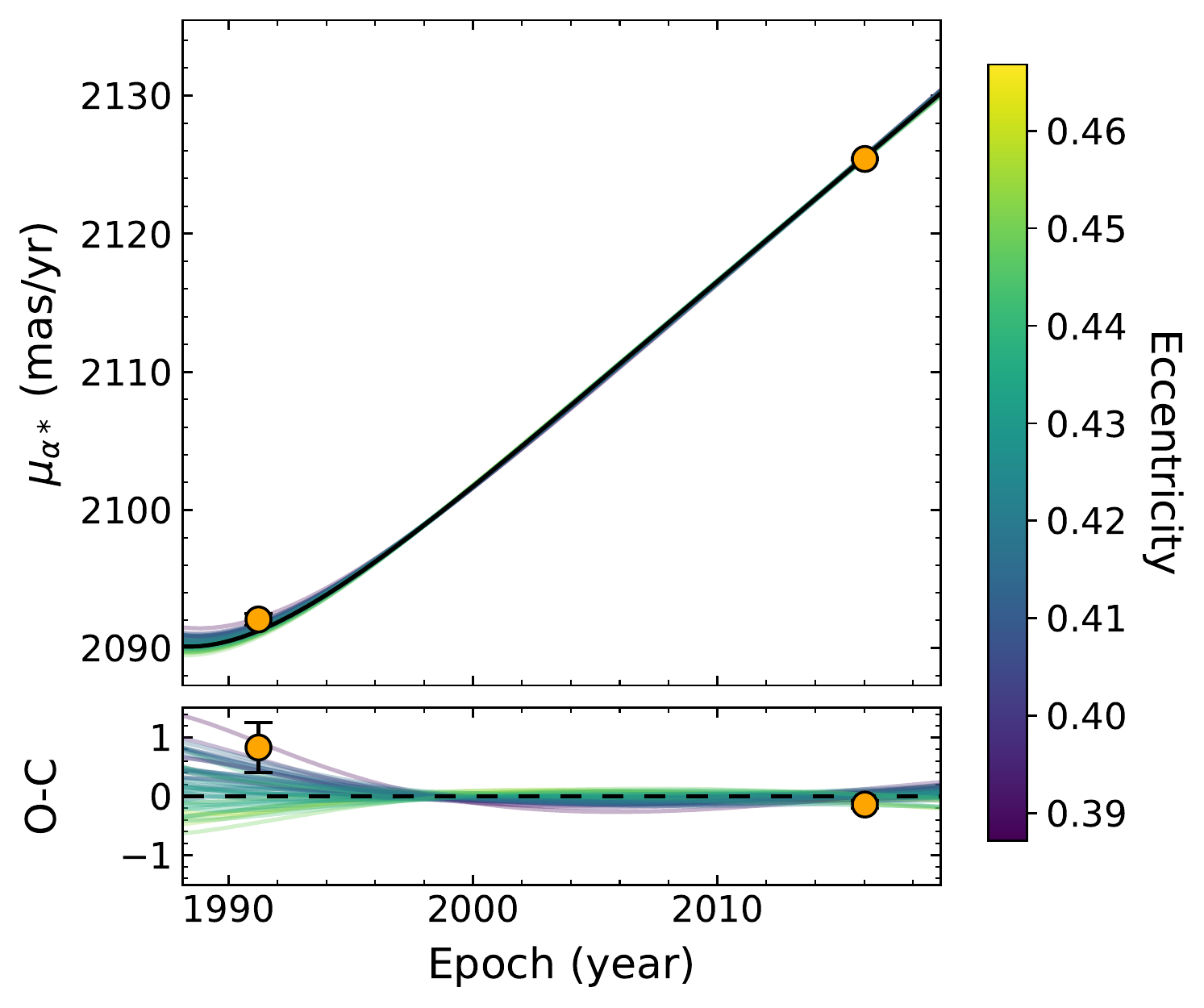}
    \includegraphics[width=.49\textwidth]{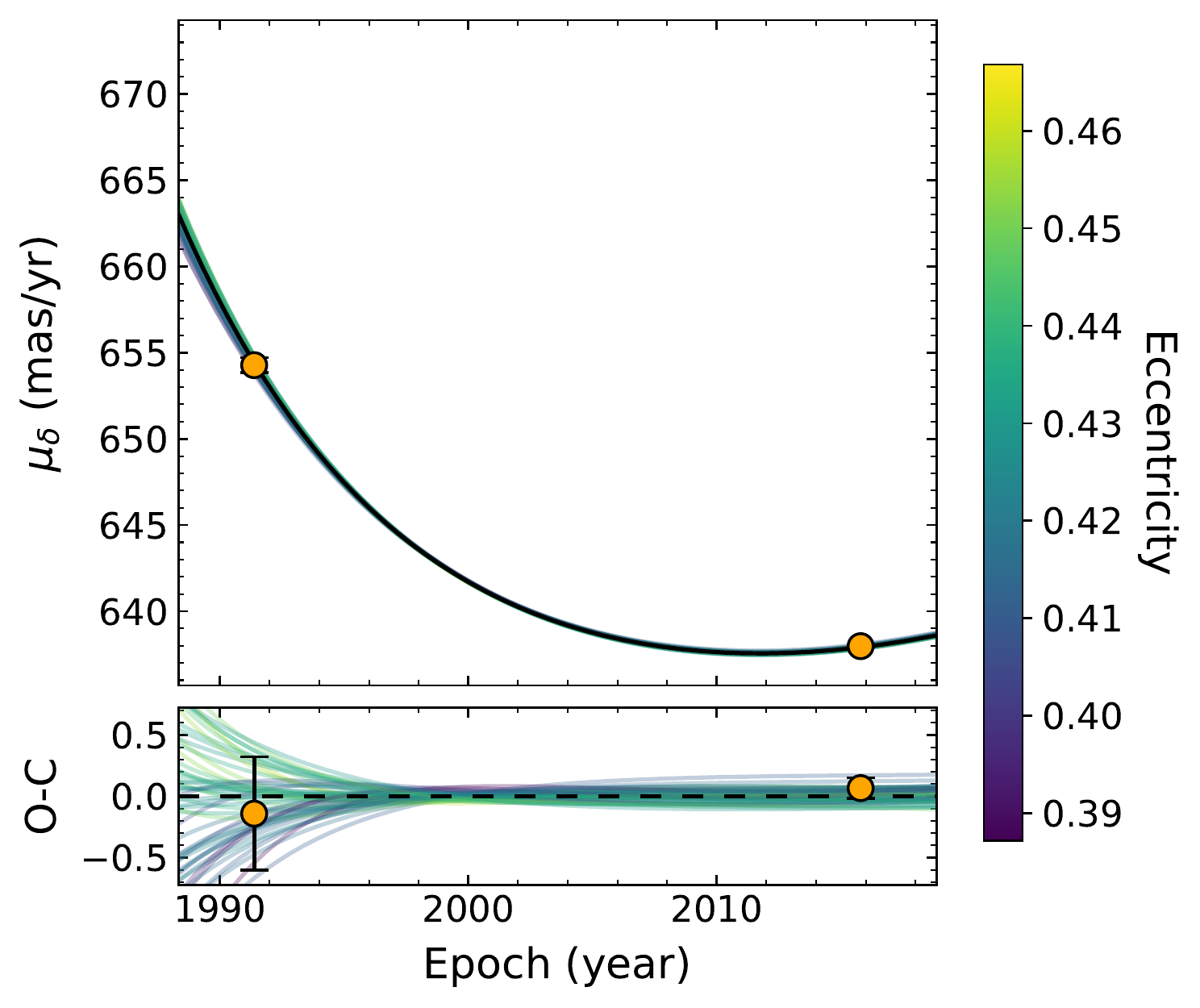}
    \caption{MCMC fitting results of the RV orbit (top), separation and position angle (middle), and proper motions (bottom). The black curve indicates the highest likelihood orbit; the 50 colorful curves are randomly selected from the posterior and color-coded by eccentricity. For clarity, the influence of the inner planet Gl~86~Ab has been subtracted from all panels. Data points in all figures are shown with filled circles, while the small lower panels show the residuals of the fit.}
    \label{fig: 6_fitting_results}
\end{figure*}

Figure \ref{fig: astrometric_B} shows the relative astrometric orbit of Gl~86~AB, and Figure \ref{fig: 6_fitting_results} shows the radial velocity, separation, position angle, and proper motions as a function of time. For display purposes, we have removed the signal from the planet Gl~86~Ab, as it has a very short period and would obscure the overall trend.

\section{The primordial orbit of Gl~86} \label{sec: primordial orbit}

In the previous section we obtained the orbital parameters of the current Gl~86 system. Next, we work out the primordial orbit when both stars were on the main sequence. To account for the alteration of the orbit throughout the period of Gl~86~B's mass loss, a simulation is necessary. Unfortunately, most stellar evolution codes, such as MESA \citep{Paxton2011}, cannot evolve stars backward in time. So, we set up MESA with a suite of different initial masses of Gl~86~B and adopt the one whose final mass is closest to its current mass, and use Mercury \citep{Chambers2012}, a general-purpose N-body integration package, to simulate the evolution of Gl~86’s orbit in Section \ref{sec: binary evolution}. 
%We confirm that the secular evolution relations of semi-major axis and eccentricity are consistent with analytic theoretical expectations.

From MESA, we find that when $M_{\rm initial} = 1.39\,\Msun$, the final mass of Gl~86~B, $0.543\,\Msun$, is closest to the mass obtained in the previous section. This initial mass is consistent with \citet{Kalirai2008}, who formulated the relation by studying the spectroscopic observations of a sample of 22 white dwarfs in two older open clusters, NGC 7789 and NGC 6819, plus data from the very old cluster NGC 6791, measuring the current masses of those white dwarfs, and calculating their corresponding progenitor masses. Their relation,
\begin{equation}
M_{\rm final} = (0.109 \pm 0.007) M_{\rm initial} + 0.394 \pm 0.025\, \Msun.
\label{eq: initial-final mass relation1} 
\end{equation}
is good down to $M_{\rm initial} = 1.16\,\Msun$.  With this relation, the initial mass of Gl~86~B is predicted to be $1.36\pm0.25\,\Msun$, which is consistent with our MESA result.  We adopt $M_{\rm B,\,initial} = 1.39\,M_\odot$ throughout this section.

In the remainder of this section, we first review analytic theoretical expectations for the evolution of semimajor axis and eccentricity and confirm that they are consistent with Mercury's results.  We then evolve the orbit of a 1.39\,$M_\odot$ star around a \MAnoerr star to infer the initial dynamical configuration of Gl~86~AB.

\subsection{Mass loss} \label{subsec: mass loss}

After a star evolves off the main sequence and passes the sub-giant branch, its radius expands and its envelope becomes loosely bound. The radiation pressure due to photon flux expels the envelope. The star will experience mass loss during the red giant branch (RGB, $\dot{M} \approx -10^{-8}\,\Msunperyr$), the asymptotic giant branch (AGB, $\dot{M} \approx -10^{-8}~ \text{to} -10^{-4}\,\Msunperyr$), and planetary nebula ($\dot{M} \approx -10^{-6}\,\Msunperyr$) phases, during which mass is cast away in the form of an isotropic stellar wind.  

\begin{figure*}
  %\centering
  \includegraphics[width=0.48\textwidth]{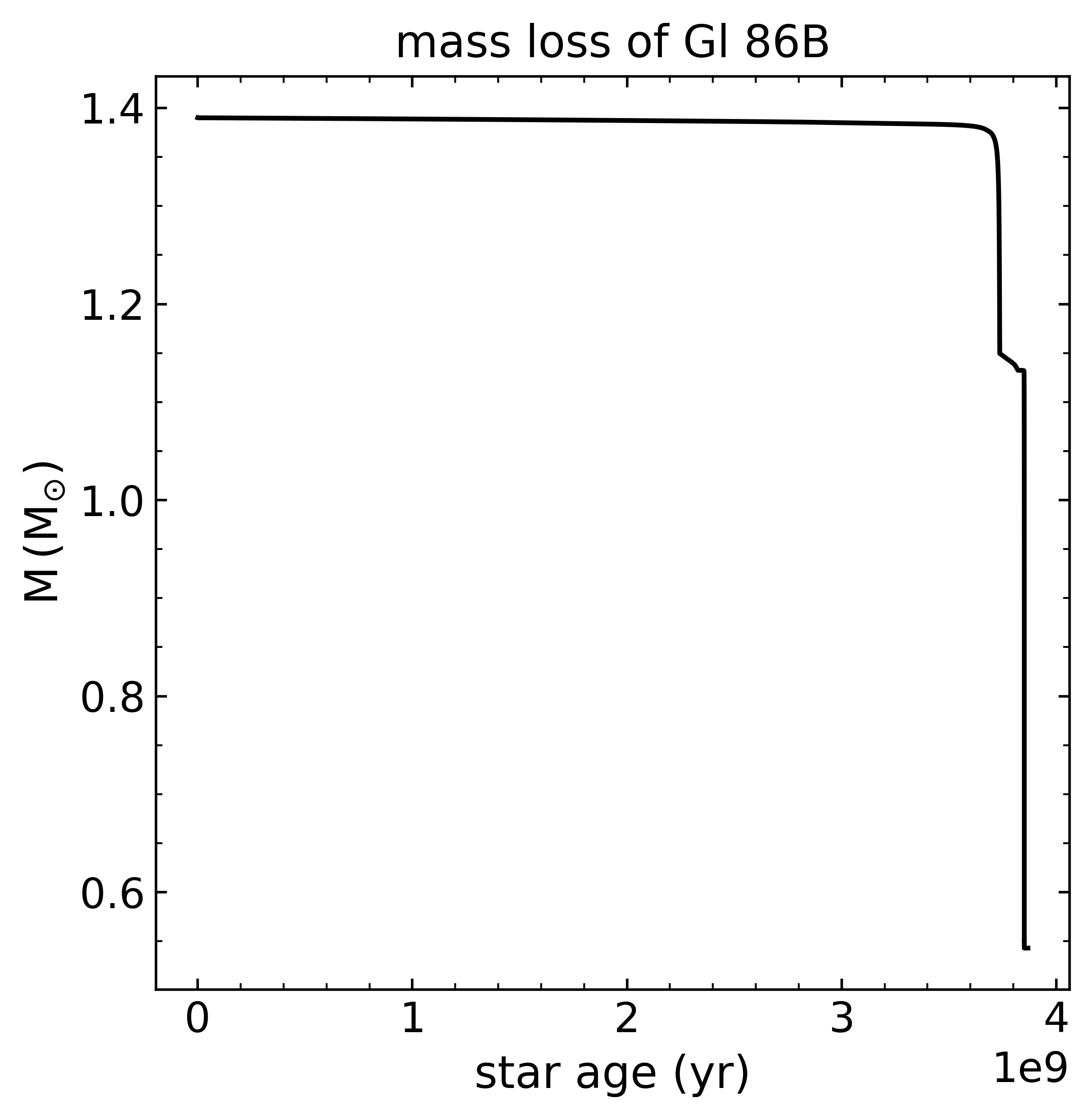}
  \includegraphics[width=0.48\textwidth]{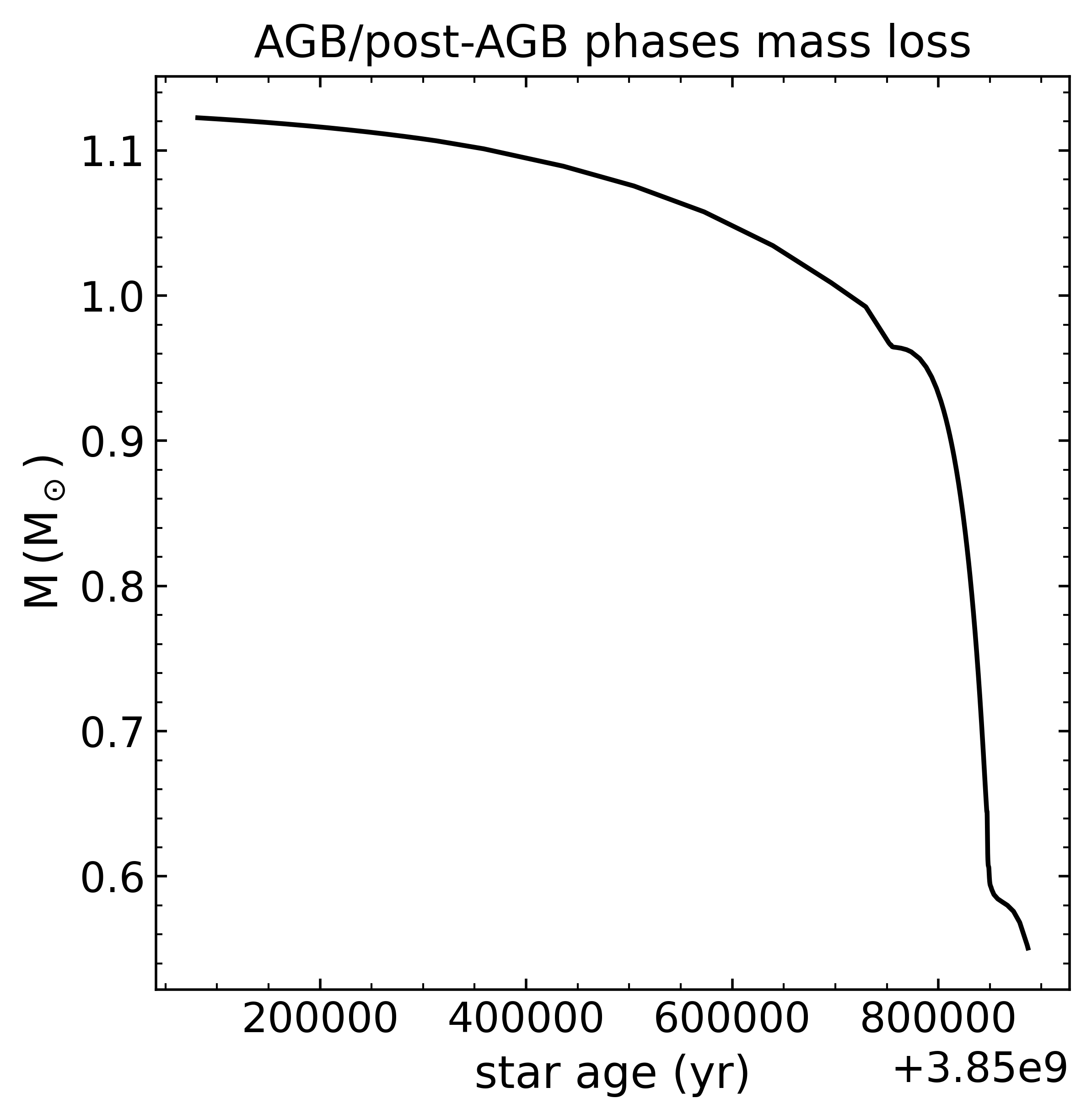}
  \caption{\textit{Left}: The mass of Gl~86~B versus star age. \textit{Right:} The mass loss of the Gl~86~B was most rapid during the AGB and post-AGB phases, where it lost $0.478\, \Msun$ in $0.602$ million years.  However, even during the final stages, mass was lost over several $10^4$\,yr, much longer than the $\approx$100\,yr orbital time of Gl~86~B.}
  \label{fig: m_Gl86B vs t}
\end{figure*}

MESA computes the (isotropic) mass loss of Gl~86~B as a function of time. Figure \ref{fig: m_Gl86B vs t} shows the mass of Gl~86~B versus the star age. Although Gl~86~B lost mass most rapidly by the end of the AGB, it lost $0.478\,\Msun$ in $0.602$ million years, equivalent to a rate of $7.95 \times 10^{-7} \,\Msunperyr$, if we approximate the average mass loss as linear. Even during the final thermal pulses that expel the envelope, mass loss proceeds on a timescale of $\sim$10$^4$\,yr.  This remains much longer than the orbital period of $\approx$100\,yr, making the mass loss very nearly adiabatic.

We can apply this in the relation between speed $v$ and separation $r$ for a Keplerian orbit with semimajor axis $a$, 
\begin{equation}
    v = 2GM_{\rm tot} \left(\frac{1}{r} - \frac{1}{2a}\right)
    \label{eq: v(r)}.
\end{equation}
We differentiate with respect to time, setting $\dot{v} = 0$ for isotropic mass loss, and take the orbit average: 
\begin{equation}
    -\frac{1}{M_{\rm tot}}\dv{M_{\rm tot}}{t} = \frac{1}{a(\left\langle \frac{2a}{r} \right\rangle - 1)} \dv{a}{t},
    \label{eq: dif v(r)}
\end{equation}
where
\begin{equation}
    \left\langle \frac{2a}{r} \right\rangle = \frac{1}{P} \int_{0}^{P} \frac{2a}{r(t)} dt = 2,
    \label{eq: ave 2r/a}
\end{equation}
and $P$ is the orbital period. Equations \eqref{eq: ave 2r/a} and \eqref{eq: dif v(r)} yield
\begin{equation}
    M_{\rm tot}a = \textrm{constant},
    \label{eq: ma=cons}
\end{equation}
which recovers Equation (16) of \citet{Jeans1924}.

For the secular evolution of the eccentricity, \citet{Hadjidemetriou1963} presents the evolution of eccentricity in a binary system with slow mass loss,
\begin{equation}
    \dv{e}{t} = -(e + \cos f)\frac{\dot{M}}{M},
    \label{eq: de/dt}
\end{equation}
where $f$ is the true anomaly. Taking the time average and using Equation (13) of \citet{Dosopoulou2016II}, the secular result is then 
\begin{align*}
    \left\langle \dv{e}{t} \right\rangle &= \frac{1}{P} \int_{0}^{P} \dv{e}{t} dt \\
    &= \frac{1}{P} \int_{0}^{P} -\frac{e + \textrm{cos}f}{M} \dv{M}{t} dt  \\
    &= -\frac{\dot{M}(1 - e^2)^{3/2}}{2\pi M} \int_{0}^{2\pi} \frac{e+\textrm{cos}f}{(1+e\textrm{cos}f)^2} df \\
    &= 0. \numberthis
    \label{eq: ave de/dt = 0}
\end{align*}

Equations \eqref{eq: ma=cons} and \eqref{eq: ave de/dt = 0} show that isotropic, adiabatic mass loss will expand the orbit, but at fixed eccentricity. This still holds with anisotropic mass loss, as long as we keep the adiabatic assumption and further assume that the anisotropic wind and jets are symmetric with respect to the equator \citep{Veras2013,Dosopoulou2016II}. 

\begin{figure}
    \centering
    \includegraphics[width=\linewidth]{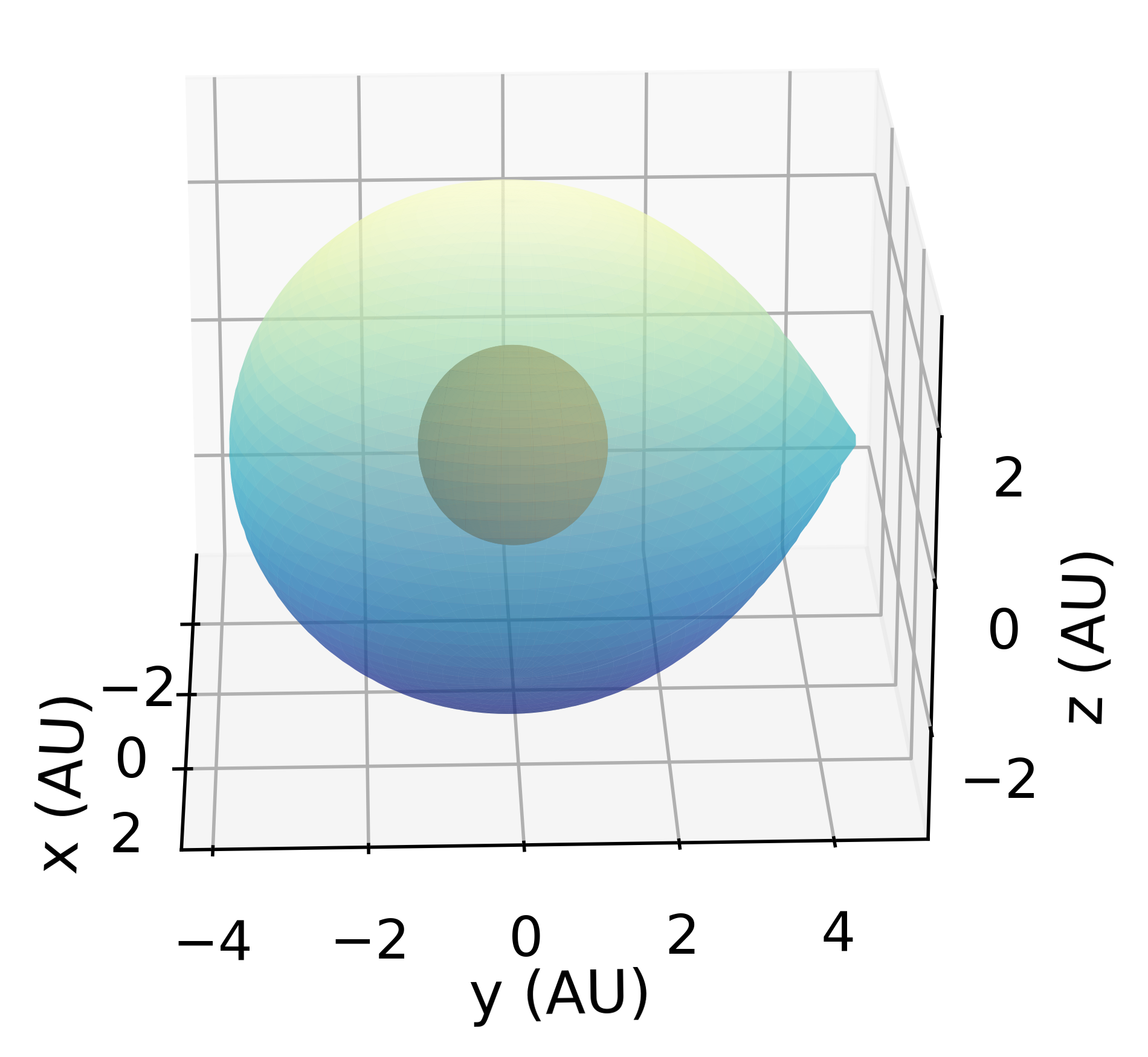}
    \caption{The blue and yellow shade is the 3D Roche lobe of Gl~86~B when it was in the post-AGB phase, during which Gl~86~B expanded most. The orange sphere represents the size of Gl~86~B at that time. The figure demonstrates that it is safe to assume there is no mass transfer, as Gl~86~B is far from touching the surface of the Roche lobe.}
    \label{fig: Roche_lobe}
\end{figure}

We next consider mass transfer due to the ejection of Gl~86~B's envelope and its possible accretion onto Gl~86~A through the Roche lobe. Once a star is large enough to fill out its Roche lobe, it transfers mass to its companion via the first Lagrangian point $L_1$. We can tell whether a star passes its Roche lobe or not using a byproduct of the MESA simulation, the radius of Gl~86~B as a function of time, and a calculator for Roche lobe properties \citep{Leahy2015}, to show the 3D illustration of the Roche lobe of Gl~86~B (Figure \ref{fig: Roche_lobe}). The result shows that the Roche lobe is very spacious, and the surface of Gl~86~B is far from touching that of the Roche lobe when Gl~86~B reaches its maximum size during the AGB stage. This suggests that there was no mass transfer/ejection throughout the evolution of the binary, when the primordial semi-major axis was \aprimordial (see Section \ref{sec: binary evolution}). 

\subsection{Binary evolution} \label{sec: binary evolution}
In Section \ref{subsec: mass loss}, we have argued that Gl~86~B's mass loss was mostly isotropic. We further assume that the mass loss was adiabatic and linear for the whole duration when Gl~86~B lost mass, because the fractional mass loss rate is small on an orbital time scale.  Our simplifying assumption of a constant mass loss rate does not change the evolution's overall physical results (see Equation \eqref{eq: ma=cons}). 

\begin{figure}%[htp]
    \centering
    \includegraphics[width=\linewidth]{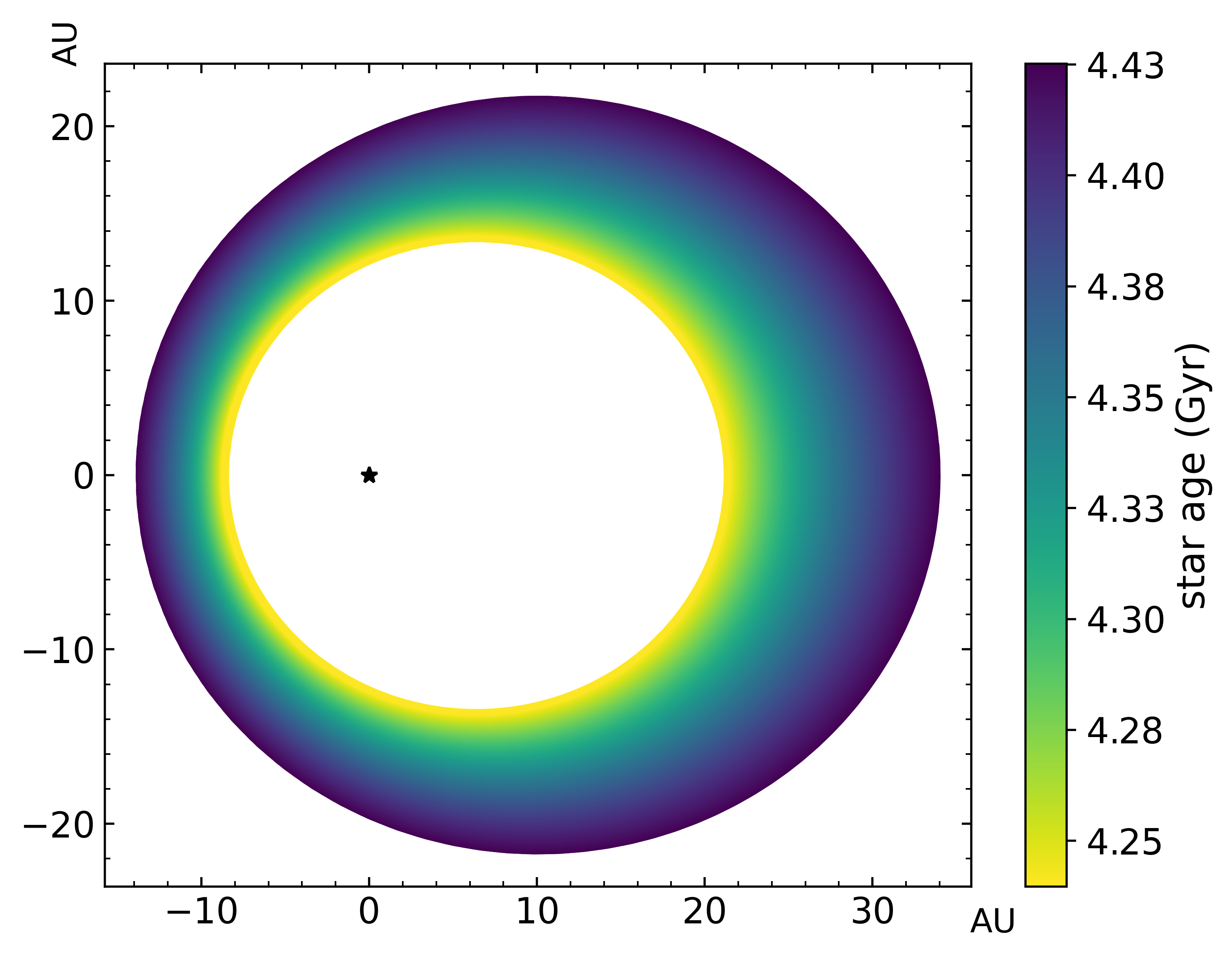}
    \caption{Mercury simulation of the orbit of Gl~86~B about Gl~86~A during the period of Gl~86~B's mass loss. The star symbol at the origin stands for Gl~86~A. When both stars were on the AGB phase, where Gl~86~B mainly started to lose mass, it was at the innermost (yellow) orbit. As it shed mass, the semi-major axis increased, and it gradually spiraled out and ended up as a white dwarf at the outermost (purple) orbit.}
    \label{fig: Gl86B orbit}
\end{figure}

\begin{figure*}%[htp]
    \centering
    \includegraphics[width=0.45\textwidth]{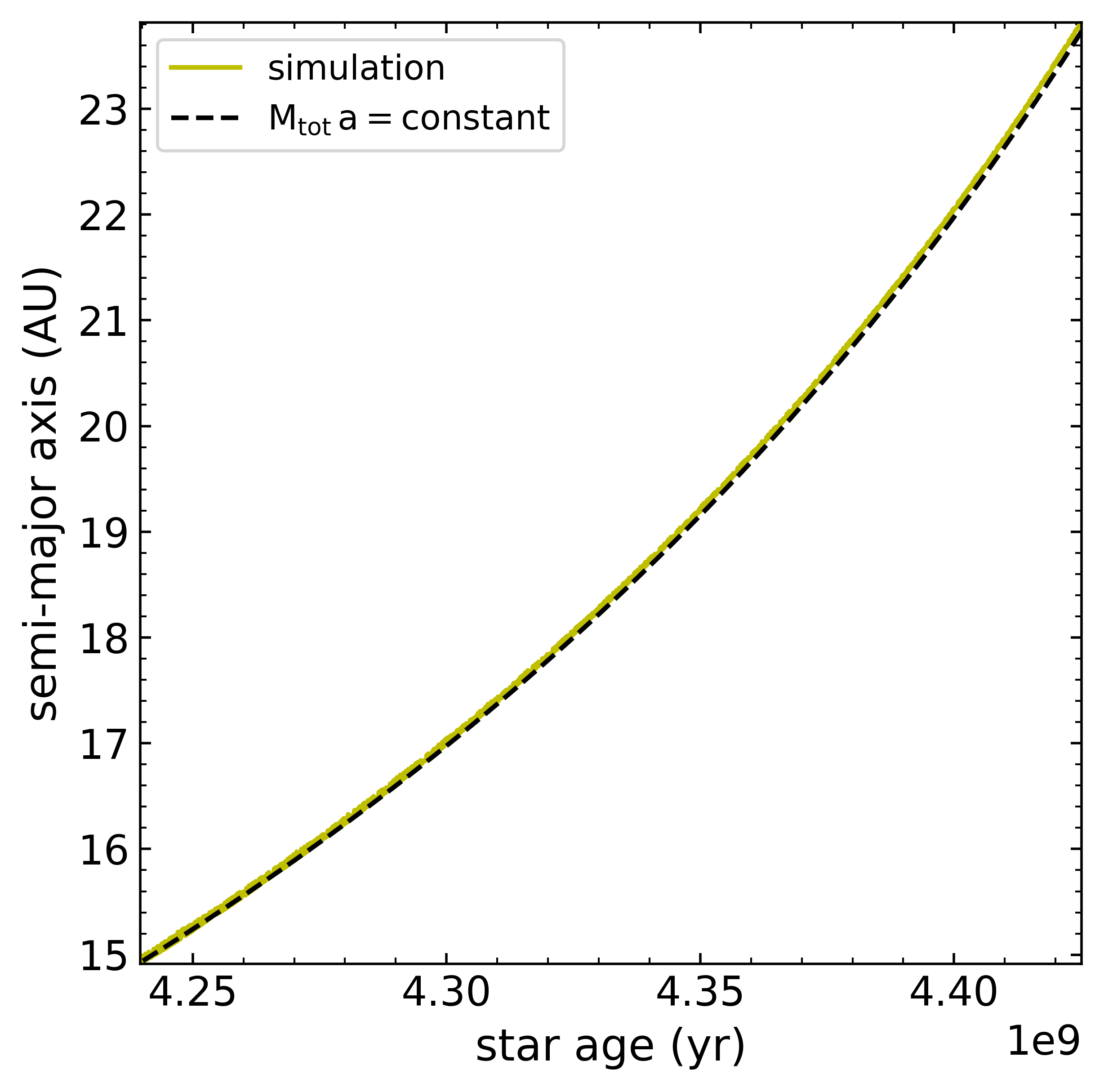}
    \includegraphics[width=0.465\textwidth]{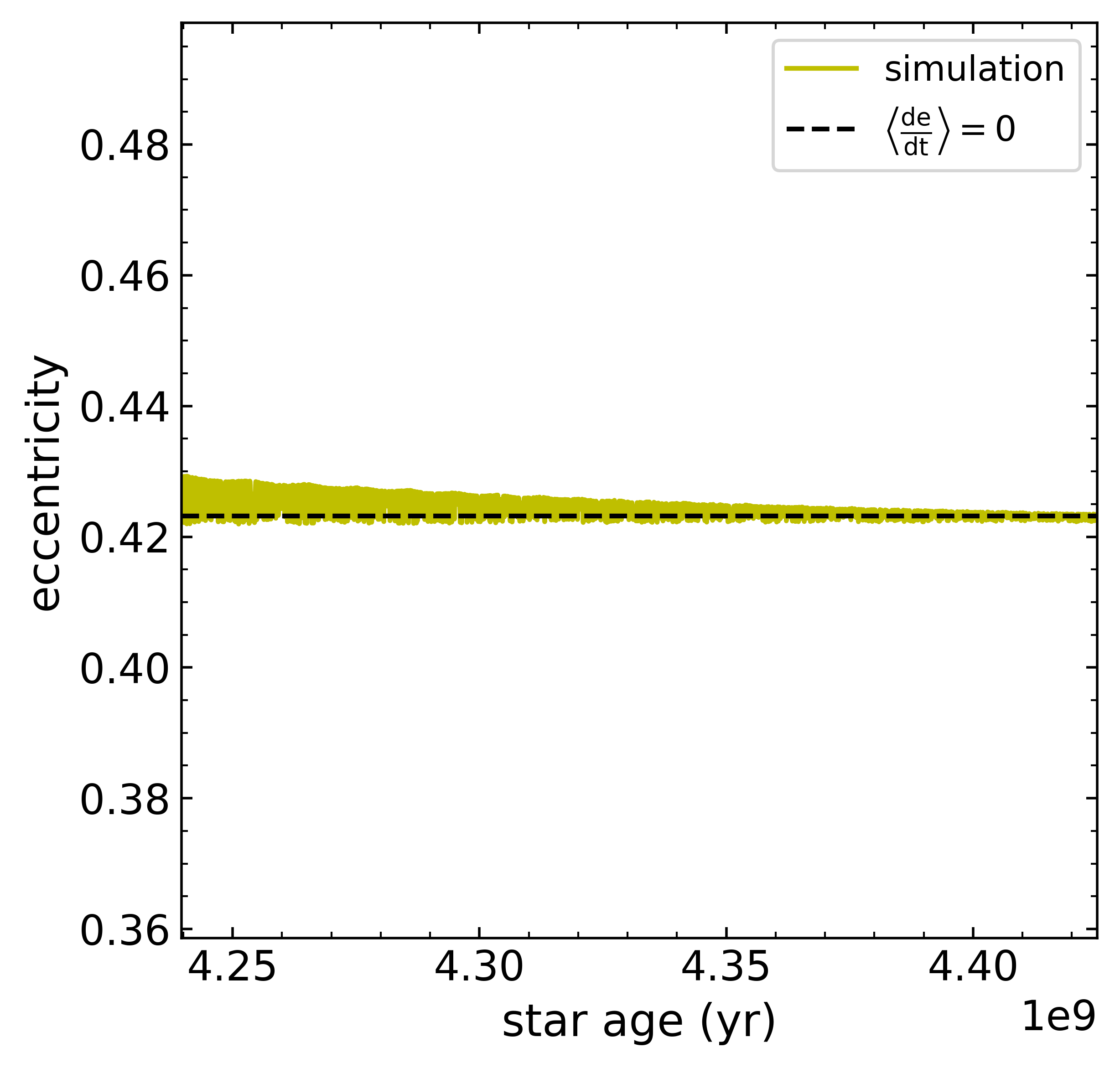}
    \caption{The secular evolution of the semi-major axis (left) and eccentricity (right) of Gl~86’s orbit since the AGB phase of Gl~86~B assuming constant mass loss. The yellow curves are the simulation results from Mercury. The dashed black curves are analytic expectations from theory. The semi-major axis goes from \aAGB to \anownoerror and the eccentricity remains unchanged.}
\label{fig: evo a and e}
\end{figure*}

We integrate the orbit of Gl~86~B around Gl~86~A using Mercury with the general Bulirsch-Stoer algorithm. It is slow but accurate in most situations and can dynamically adjust the step size. Our Mercury integration reproduces the orbit of the companion with respect to the host (Figure \ref{fig: Gl86B orbit}) and enables us to visualize its slow and steady expansion throughout time.
 Mercury's simulation agrees with the secular evolution of semi-major axis and eccentricity in Equation \eqref{eq: ma=cons} and Equation \eqref{eq: ave de/dt = 0}; these are shown in the two panels of Figure \ref{fig: evo a and e}.

\section{Feasibility and challenges of planet formation in the Gl~86 system} \label{sec:formation}

\subsection{Mechanism of planet formation} \label{sec: planet formation}

There are two main phases of the formation of a Jovian planet by core accretion. The first phase is the formation of the planetary core, for which two main theoretical mechanisms have been advanced: planetesimal accretion and pebble accretion. Planetesimal accretion, once the dominant hypothesis \citep{Pollack1996}, has difficulty forming a core and accreting a gaseous envelope before the protoplanetary disk dissipates \citep{Lambrechts-Johansen2012}. The theory of pebble accretion has become increasingly popular as a solution to this problem \citep{Bitsch201510, Johansen-Lambrechts2017, Bitsch2019}. Although planetesimal accretion still dominates during the early stage of the growth of the planetary embryo, pebble accretion becomes significant when the core is several hundred kilometers in size. Pebbles passing by the core experience the drag force of the gas in the protoplanetary disk, undergo Bondi-Hill accretion, and quickly spiral in \citep{Johansen-Lambrechts2017}.  This drastically reduces the time scale and enables the formation of the planet within the lifetime of the protoplanetary disk. 

The second phase starts after the core reaches the isolation mass, during which the core carves a shallow gap and stops the drift of pebbles from accumulating to the core. Then, the core starts to accrete from the gaseous envelope. This process is slow until runaway gas contraction initiates when the mass of the envelope exceeds that of the core \citep{Bitsch201510, Bitsch2018}.
% \li{It may be easier to directly start with the next paragraph, starting with things like "planetary formation around close binaries have recently been investigated by... "} \Zeng{Done}

In rare cases, a planet could start as a circumbinary planet and be tidally captured by one of the stars, ending up as a circumstellar planet \citep{Gong2018}. It can also form around one star and be captured by the other \citep{Kratter-Perets2012}.
However, both mechanisms typically produce highly eccentric planets, in conflict with Gl~86~Ab's low observed eccentricity of \eplanet.  It is possible that Gl~86~Ab was captured as an eccentric planet and tidally circularized.  However, its orbital period is 15.8 days, while the tidal circularization period cutoff for the 4~Gyr-old open cluster M67 has been measured to be $\approx$12 days \citep{Meibom..Mathieu..2005,Geller..Mathieu..2021}.  Planets on longer periods remain eccentric; they need more than 4~Gyr to circularize.  The cooling age of the white dwarf Gl~86~B is just $1.25 \pm 0.05$~Gyr \citep{Farihi2013}, far too little time for tidal circularization of Gl~86~Ab.  In the rest of the paper, we consider only the scenario in which Gl~86~Ab formed around its current host star.

%While such scenarios exist, the chances are only around 10\%, based on extensive numerical simulations. We ignore such a possibility in this paper.

The mechanism of planetary formation in specific close binaries has been investigated by \citet{Jang-Condell2015}. They studied the feasibility of planet formation in HD~188753~A, $\gamma$~Cep~A, HD~41004~A, HD~41004~B, HD~196885~A, and $\alpha$~Cen~B. They tested a suite of eighteen different combinations of viscosity parameters $\alpha$ and accretion rates $\dot{M}$, given the binary parameters $M_{\rm host}$, $M_{\rm comp}$, $a$, and $e$, and counted how many ($\alpha$, $\dot{M}$) models satisfied core accretion or disk instability mechanism, respectively. The conventional criterion for core accretion to occur is whether the total solid mass in the disk exceeds $10 \, \Mearth$, the least amount of mass to create a rocky core and initiate gas accretion. \cite{Jang-Condell2015} found that except for HD~188753~A, where none of the models fit the observed system properties, core accretion was overwhelmingly more likely than disk instability for the formation of the planet. \citet{Jang-Condell2008} also pointed out that disk instability can only occur in the most massive disks with extremely high accretion rates. 

In order to form Gl~86~Ab, we must then satisfy two requirements.  First, we must have enough solid material in the disk to assemble a $\approx$10\,$M_\oplus$ core.  Second, the disk itself must exceed the current mass of Gl~86~Ab in order to supply the gaseous envelope.  In the following section we will work out the total mass of Gl~86's disk under different models.  We will assume a minimum dust mass of 10\,$M_\oplus$ and a minimum total mass of 5\,$M_{\rm Jup}$ in order to have any chance of forming Gl~86~Ab. 

\subsection{Total disk mass and dust mass}

In this section we compute both the total mass of the disk and its mass in dust suitable for forming the core of Gl~86~Ab.
% We trace dust in the protoplanetary disk around Gl~86~A by assuming that it follows the gas.
We take the total dust mass to be a dust-to-gas ratio multiplied by the total disk mass, which is the integral of the disk's surface density from the inner rim to the outer truncation radius. We will compute the truncation radius due to the stellar companion first and the inner rim next. 

\subsubsection{Truncation radius}

The protoplanetary disk around Gl~86~A could not have extended past the orbit of Gl~86~B, and in fact would have been truncated at a significantly smaller radius.  The first criterion that needs to be satisfied to truncate a disk is that the resonant torque should be greater than the viscous stresses. A viscosity-dependent tidal distortion and resonant interactions exert torques on the disk with opposite directions. If the torque of the resonant interaction surpasses that of the tidal distortion, a gap would be opened and the disk would be truncated. 

The magnitude of resonant torques varies at different resonant states $(m, l)$, where $m \ge 0$ is the azimuthal number, and $l$ the time-harmonic number. At a given resonant state, there are three different types of resonances: the inner Lindblad resonance (ILR), the outer Lindblad resonance (OLR), and corotational resonance (CR). They occur at different positions according to Equations (9) and (10) of \cite[][hereafter AL94]{Artymowicz-Lubow1994}:
\begin{gather}
 r_{\rm CR} = (m/l)^{2/3} \mustar^{1/3} a 
 \label{eq: r_CR} \\
 r_{\rm LR} = [(m \mp 1)/l]^{2/3} \mustar^{1/3} a
 \label{eq: r_LR}
\end{gather}
where $\mustar$ equals $M_{\rm host}/(M_{\rm host} + M_{\rm comp})$ for the circumstellar disk of $M_{\rm host}$, and $M_{\rm comp}/(M_{\rm host} + M_{\rm comp})$ for the circumstellar disk of $M_{\rm comp}$, and $1$ for the circumbinary disk. The minus and plus signs in Equation \eqref{eq: r_LR} correspond to the inner and outer LR, respectively. The ILR dominates in the circumstellar case, and OLR dominates in the circumbinary case. We are looking for the smallest possible radius at which the resonant torque is greater than the viscous stress or satisfies Equation (15) of \citetalias{Artymowicz-Lubow1994}. We check the ILR first because it dominates in a circumstellar disk, and it has the smallest radius among the three types of resonant interactions given a $(m, l)$ set.  With the method elaborated in \citetalias{Artymowicz-Lubow1994}, this first criterion breaks down as
\begin{equation}
    \alpha^{1/2} \left(\frac{H}{r}\right) = \left(\frac{a\left|\phi_{\rm ml}\right|}{GM}\right) \frac{(\pi m)^{1/2}(m \pm 1)^{1/6}\left|\lambda \mp 2m\right|}{2\mustar^{2/3}l^{2/3}},
    \label{eq: eq16}
\end{equation}
where $\lambda = m$ for the ILR of the circumstellar disk and $\lambda = -(m + 1)$ for the OLR of the circumbinary disk, and $\alpha$ is the viscosity parameter \citep{Shakura-Sunyaev1973}. The LHS of Equation \eqref{eq: eq16} indicates the magnitude of viscous stress. The larger the $\alpha$, the stronger the viscous stress. We can also write the LHS as $\frac{1}{\sqrt{{\rm Re}}}$ in terms of Reynolds number ${\rm Re}$, where
\begin{equation}
    {\rm Re} = \frac{1}{\alpha}\left(\frac{r}{H}\right)^2.
    \label{eq: Reynolds}
\end{equation}

\begin{figure*}[t!]
    \centering
    \includegraphics[width=.5\linewidth]{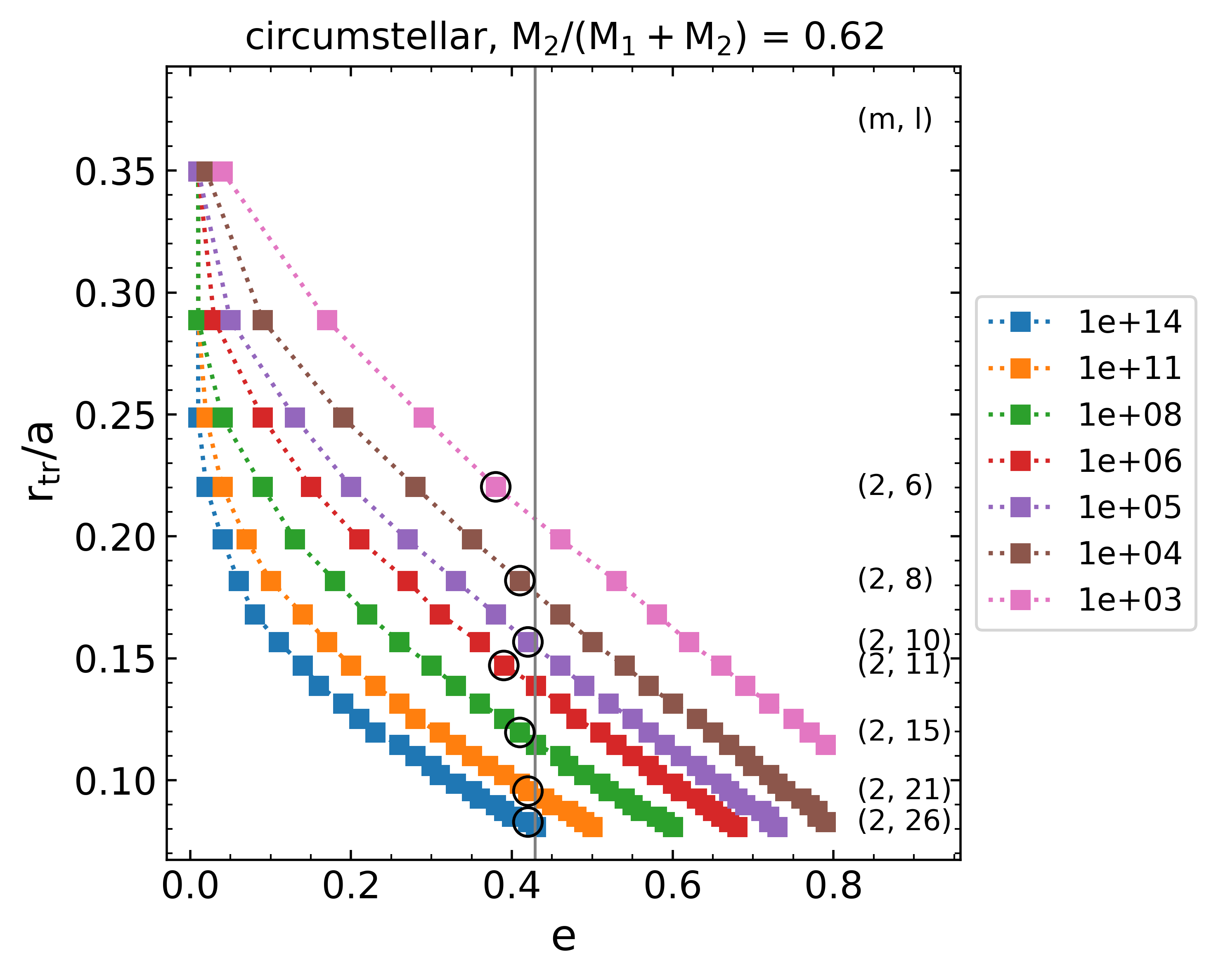}
    % \vspace{-30mm}
    \includegraphics[width=.4\linewidth]{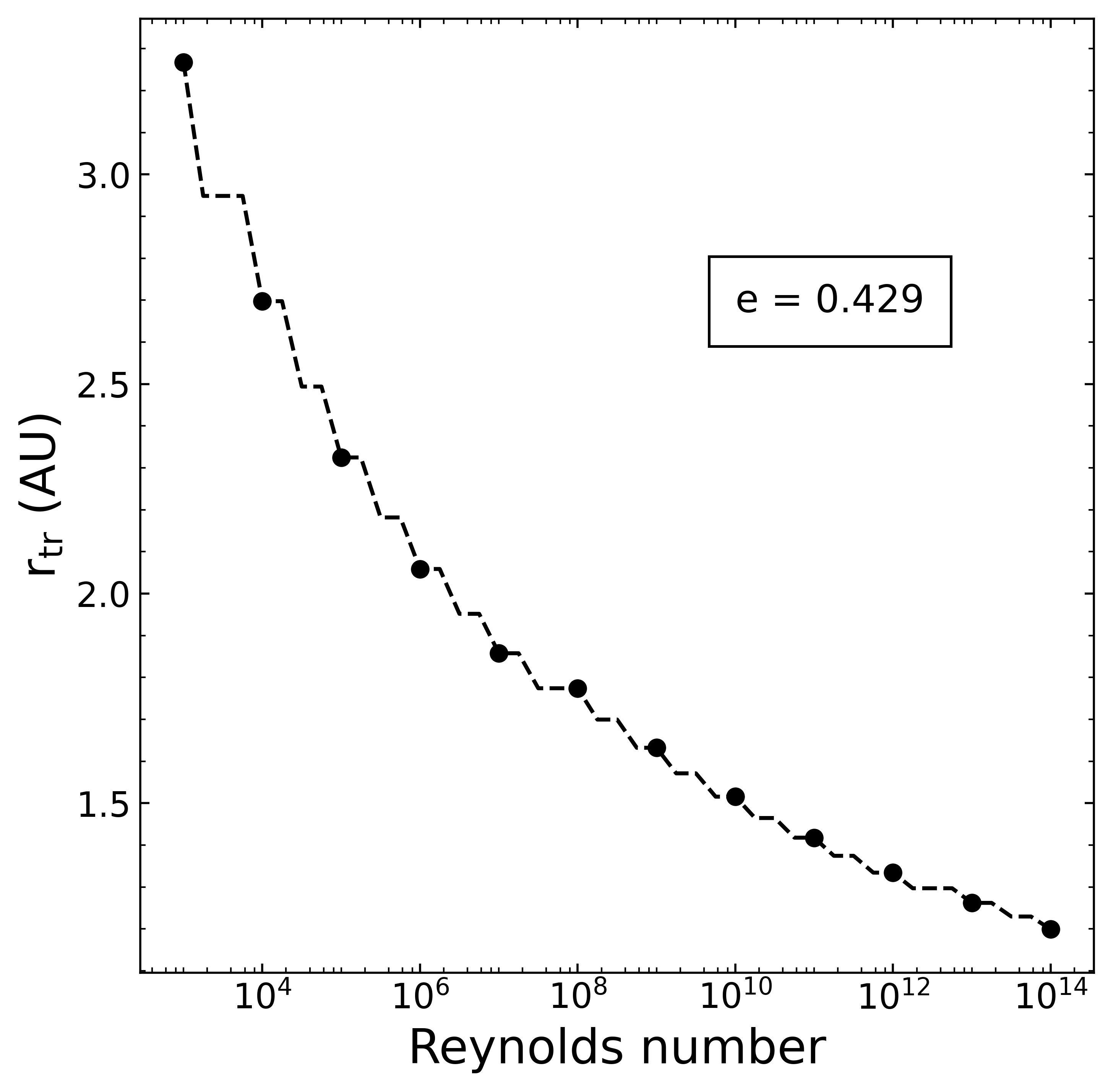}
    \caption{\textit{Left}: Locations of resonant interactions between the circumstellar disk around the present Gl~86~A and the primordial orbit of Gl~86~A and B.  The present eccentricity is shown by a vertical line, while different colors indicate different Reynolds numbers. Because the Reynolds number is inversely proportional to the viscosity coefficient, a larger Reynolds number means smaller viscous stress. The squares on each strip represent a large enough eccentricity that gives rise to a resonant torque that overcomes the viscous stress. %The grey line indicates the eccentricity of the Gl~86 binary system. 
    From the plot, we can read off the truncation radii of Gl~86's protoplanetary disk for different Reynolds numbers. Black circles indicate those radii \citepalias{Artymowicz-Lubow1994}. 
    \textit{Right}: The truncation radii in terms of the Reynolds numbers at $e =$ \enoerr. The black dots are the truncation radii corresponding to a sequence of Reynolds numbers starting from $10^3$ to $10^{14}$, incrementing with a factor of $10$ for each step.
    }
    \label{fig: rtr_vs_e}
\end{figure*}

With $M_1 =$ \MAnoerr and $M_2 = 1.39\,\Msun$, the left panel of Figure \ref{fig: rtr_vs_e} shows the radial positions where resonant interaction occurs in units of semi-major axis versus the distribution of the eccentricity at which the torque is large enough to balance the viscous stress. As the eccentricity increases, the magnitude of the resonant torque gets larger, which means the disk can be truncated more easily. Squares connected by dotted lines share the same Reynolds number. The grey line in the figure indicates that the eccentricity of Gl~86 is \enoerr. We can read the positions where the circumstellar disk is truncated, based on how the grey line intersects with the $r$ vs.~$e$ curves. For Reynolds numbers equal to $10^3, 10^4, 10^5, 10^6, 10^8, 10^{11}, 10^{14}$, the truncation radii are $0.22a, 0.18a, 0.16a, 0.15a, 0.12a, 0.10a, 0.08a$, or $3.3, 2.7, 2.3, 2.2, 1.8, 1.4, 1.2$ AU, respectively. The right panel shows the resulting truncation radii with respect to different Reynolds numbers, ranging from $10^3$ to $10^{14}$, at $e = 0.429$ and in terms of AU. The black dots are the truncation radii corresponding to a sequence of Reynolds numbers starting from $10^3$ to $10^{14}$, incrementing with a factor of $10$ for each step.

The second criterion is that gap opening time $t_{\rm open}$ should last for a reasonably short time relative to the viscous timescale. The gap opening time, approximately the same as the viscous closing time, equals $t_{\rm open} \approx (\Delta r)^2 / \nu$, where $\Delta r$ is the radial extent of the gap, and $\nu$ the viscosity. According to the formula of viscosity,
\begin{equation}
\nu = \alpha c_{\rm s} H,
\end{equation}
where $c_{\rm s}$ the sound speed, and $H$ the height of the disk. $H = c_{\rm s} / \Omega$, where $\Omega$ is the Keplerian orbital angular frequency  
\begin{equation}
\Omega = \sqrt{\frac{G(M_1 + M_2)}{r^3}}. 
\label{eq: Omega}
\end{equation}
So,
\begin{align}
\nu &= \alpha c_{\rm s} H \nonumber \\
    &= \alpha H^2 \Omega. \label{eq: nu}
\end{align}
With Equations \eqref{eq: Reynolds}, \eqref{eq: nu}, and $\Omega = \frac{2 \pi}{P}$, we have
\begin{align}
\frac{t_{\rm open}}{P} &\approx \left( \frac{\Delta r}{r} \right)^2 \frac{{\rm Re}}{2 \pi} \nonumber \\
    & = \left( \frac{\Delta r}{r} \right)^2 \frac{1}{2 \pi \alpha} \left( \frac{r}{H} \right)^2 \nonumber \\
    & \approx \frac{1}{2 \pi \alpha}.
\end{align}
Since $\alpha$ ranges between $0.001 - 0.1$, $t_{\rm open}$ is approximately $1\,P - 100\,P$, which is short compared to the disk's lifetime on the main sequence. 

\subsubsection{Inner rim}

The inner part of a protoplanetary disk can be divided into four components: a dust-free region, a dust halo, a condensation front, and an optically thick disk (Figure 1 of \citealt[][henceforth U17]{Ueda2017}). The dust in this region functions as a feedback regulation. If the temperature of the disk is lower than $T_{\rm ev}$, then the dust condenses and heats up. Otherwise, the temperature goes down when the dust evaporates since the emission-to-absorption ratio of the dust is lower than that of the gas \citepalias{Ueda2017}. So, the temperature in this region is approximately high enough to evaporate dust, and the inner rim of the disk lies between the dust halo and condensation front. According to \citet{Flock2016} and \citetalias{Ueda2017}’s simulations, based on $T_* = 10000$\,K, $M_* = 2.5 \,\Msun$, and $R_* = 2.5\,\Rsun$ models, the resulting radial profile of mid-plane temperature is step-like, with a temperature of $T_{\rm ev} \approx 1470$\,K in the dust halo region spanning from $0.35 - 0.45$\,AU from the star. 

Such stellar parameters resemble those of Herbig Ae/Be stars, pre-main-sequence stars with a mass between $2-8\,\Msun$ \citep{Natta2001}. \citet{Kama2009} studied such stars and verified observational data by exploring the inner rim of their protoplanetary dust disks with a Monte Carlo radiative transfer simulation. They worked out an analytical expression of $\Rrim$ in terms of dust, disk, and stellar properties. In terms of their result, high surface density, large silicate grains, and iron and corundum grains reduce the rim radius. With a typical surface density of $1 \,\rm g\,cm^{-2}$, ordinary $10\, \microm$ grains give a $0.4$\,AU rim, while grains as large as $100 \,\microm$ give a $0.2$\,AU rim, and those in the minor extreme $0.1 \, \microm$ give an inner rim of $2.2$\,AU. 

Gl~86~A, however, is a Sun-like star. The disk structure of its inner rim is much different, as the dust sublimation temperature is reached closer to the star. With a specific accretion rate of $3.6 \times 10^{-9}\,\Msunperyr$, irradiated hydrostatic disk models show that the silicate sublimation front begins at around $0.08$\,AU, a curved dust rim exists between $0.08$\,AU and $0.15$\,AU, a small shadowed region between $0.2 - 0.3$\,AU and a flared disk beyond $0.3$\,AU \citep{Flock2019}. The simulation also shows a steep rise of gas density at $0.13$\,AU. Therefore, we set $0.13$\,AU as the start of the inner rim of Gl~86~A’s protoplanetary disk.  This is similar to Gl~86~Ab's current orbital radius of 0.11\,AU.

\subsubsection{Disk Mass}

With the truncation radius and the inner rim worked out, the next step is to compute the total mass of the disk and the mass of solid material available to form the core of Gl~86~Ab.  In this section we derive the expression of the disk surface density and then integrate it numerically. 

The hydrostatic equilibrium equation reads
\begin{gather}
    \dot{M} = 3 \pi \nu \Sigma_{\rm G} = 3\pi\alpha H^2\Omega\Sigma_{\rm G} 
    \label{eq: hydrostatic 1} 
\end{gather}
where $\Sigma_{\rm G}$ is the surface density of gas. We combine Equations \eqref{eq: Reynolds}, \eqref{eq: Omega}, \eqref{eq: nu} and \eqref{eq: hydrostatic 1} to get
\begin{equation}
    \Sigma_{\rm G} ({\rm Re}, \dot{M}, M_1, M_2) = \frac{\dot{M}{\rm Re}}{3\pi \sqrt{G(M_1 + M_2)r}}.
    \label{eq: Sigma_G} 
\end{equation}
The integral of Equation \eqref{eq: Sigma_G} from the inner rim to the truncation radius is the total mass of a protoplanetary disk. This result is consistent with Table 1 of \citet{Jang-Condell2003}. 

\begin{figure}
    \centering
    \includegraphics[width=\linewidth]{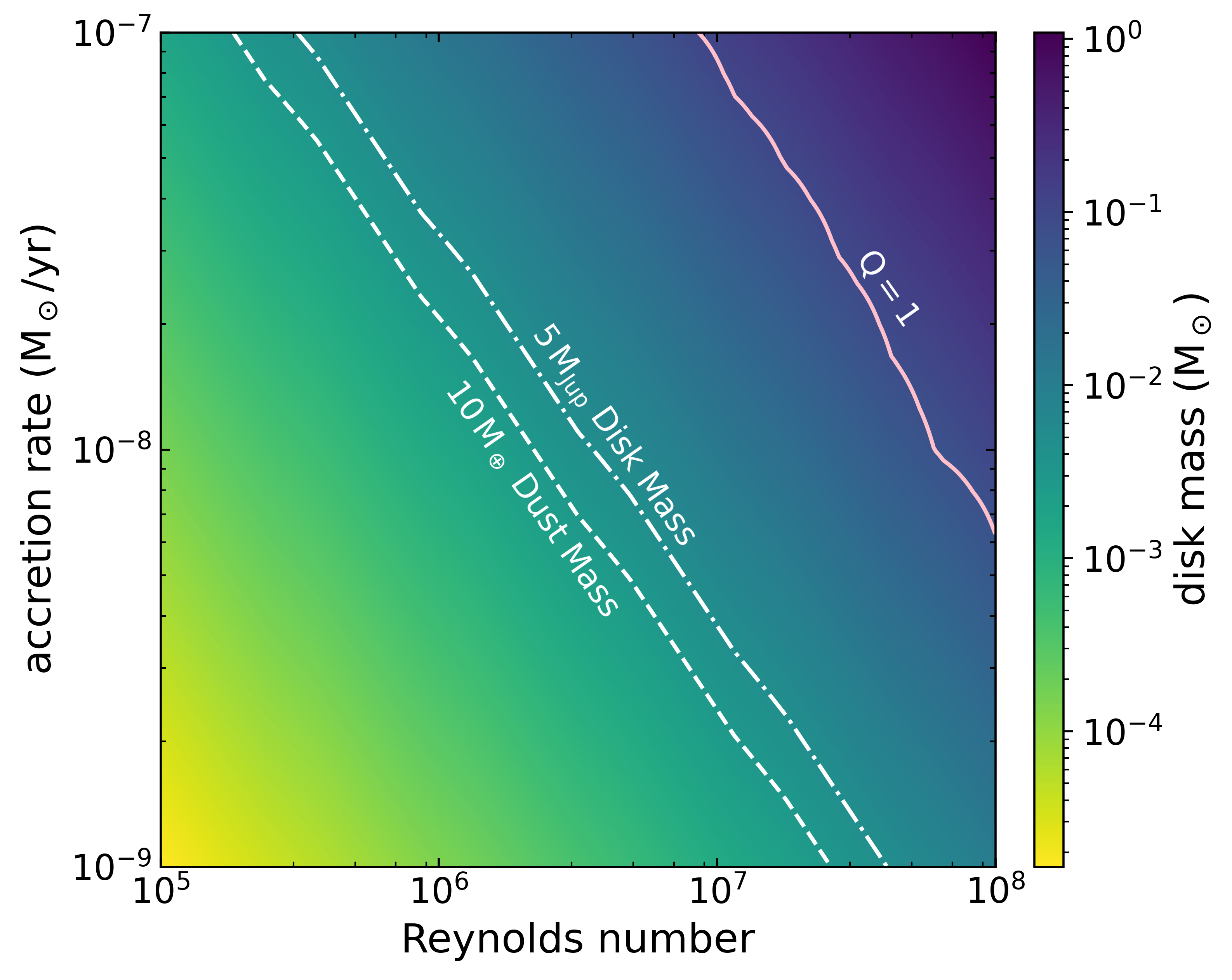}
    \caption{Total mass of the protoplanetary disk of Gl~86~A in terms of Reynolds number and accretion rate. The mass is negligible at the lower-left corner, with small Reynolds numbers and low accretion rates, and large at the upper-right corner, with large Reynolds numbers and large accretion rates. The dashed white line indicates a disk mass of $1000 \, \Mearth$, or $3.0\times 10^{-3}\,\Msun$, or a dust mass of $10 \, \Mearth$ assuming a dust-to-gas ratio of $10^{-2}$. The dash-dotted white line represents $5 \,\Mjup$ disk mass. The protoplanetary disk of Gl~86~A must have occupied the space to the right of these white lines in order to have possibly formed Gl~86~Ab.  To the right of the pink line the disk is Toomre unstable ($Q < 1$) at its truncation radius.}
    \label{fig: disk_mass_colormap}
\end{figure}

% Rework this paragraph a bit: need dust mass >10 Earth masses, need total mass >~5 Jupiter masses (current mass of Gl 86 Ab)
%In Section \ref{sec: planet formation}, we discussed that 
In order to form Gl~86~Ab, the protoplanetary disk must have had enough solids to form the planetary core and enough mass to supply the gaseous envelope.  The total dust mass is the disk mass multiplied by a dust-to-gas ratio, which starts around $10^{-10}$ near the star and increases rapidly near the inner edge of the disk. It reaches a plateau of $10^{-2}$ when it reaches the optically thick disk (e.g., Figure 2 of \citetalias{Ueda2017}). We adopt $10^{-2}$ as the dust-to-gas ratio. A minimum $10 \,\Mearth$ total dust mass implies a disk mass of 1000\,$\Mearth$, or $\approx$3\,$M_{\rm Jup}$.  The disk must also have supplied Gl~86~Ab's envelope, and must therefore have exceeded the planet's current mass of $\geq$4.3\,$M_{\rm Jup}$ (depending on the inclination).  We adopt 5\,$M_{\rm Jup}$ as the minimum disk mass that could have possibly permitted the formation of Gl~86~Ab. 

We ignore Reynolds numbers higher than $10^{8}$, as such Reynolds numbers rarely occur, and a typical Reynolds number is only $\sim$10$^5$ \citepalias{Artymowicz-Lubow1994}. For large Reynolds numbers, the truncation radii are so small that they are very close to the inner rim. However, as the disk mass is proportional to ${\rm Re}$ (see Equation \eqref{eq: Sigma_G}), it can still be unreasonably large when ${\rm Re}$ is large even if the disk mass is integrated from the inner rim to the truncation radius. Huge Reynolds numbers require extremely small viscous stresses $\alpha$, and the resulting disk mass exceeds the current mass of Gl~86~A.  

Figure \ref{fig: disk_mass_colormap} shows the total disk mass in terms of Reynolds number and accretion rate. 
With accretion rates ranging from $10^{-9} \,\Msunperyr$ to $10^{-7} \,\Msunperyr$ and Reynolds numbers from $10^5$ to $10^{8}$, the total disk mass goes from $10^{-5} \,\Msun$ to $1 \,\Msun$.  The region to the right of the dashed (dash-dotted) white line indicates dust (disk) mass higher than $10\,\Mearth$ ($5\, \Mjup$).  To the right of both lines, the disk exceeds the minimum mass to supply the material to form Gl~86~Ab. 

Figure \ref{fig: disk_mass_colormap} shows that with a Reynolds number $\gtrsim$10$^7$, the protoplanetary disk can contain enough material to form Gl~86~Ab despite its truncation at $\approx$2\,AU. This minimum mass corresponds to about twice the minimum mass Solar nebula \citep{Hayashi_1981} integrated from 0.13\,AU to 2\,AU.  Still, this requires a very high efficiency in converting disk mass to planet mass.  At very high disk masses, which would allow for a less efficient conversion of disk mass to planet mass, the disk approaches the Toomre stability limit \citep{Safronov_1960,Toomre_1964}.  The pink line on Figure \ref{fig: disk_mass_colormap} indicates that this stability limit has been breached at the disk's outer truncation radius.

Changes to our assumptions can present further difficulties in accounting for Gl~86~Ab.  One example is our assumption of a $Z = 0.010$ metallicity for the system.  Literature measurements of [Fe/H] for Gl~86~A vary from $-0.30$ \citep{Ramirez2007} to $-0.16$ \citep{Prieto2004}, while the majority fall between $-0.25$ and $-0.20$ (e.g. \citet{Santos2000, Santos2004, Ramirez2013}). We adopt $-0.23$ for [Fe/H] and $0.0134$ for solar metallicity \citep{Grevesse2012}, which give $Z = 0.0134\times 10^{-0.23} \approx 0.008$. We approximate this value to be $0.01$.  Using $Z = 0.01$ and Figure 4 of \citet{Bitsch201510}, which displays the relation between the initial formation distance, formation time, final distance, and final mass of a planet, if Gl~86~Ab is formed at $3$\,AU away from the host star (just beyond the truncation radius) and the final distance is $\sim$0.1\,AU, the figure indicates that the final mass is only several hundred times the Earth's mass. This number is a factor of $\approx$5 smaller than the mass of Gl~86~Ab. 

Another challenge is the reliability of the planet formation mechanisms, namely pebble accretion and planetesimal accretion. These mechanisms can be characterized by inefficient coagulation of the solid material onto the planetary core \citep{Guillot2014}. Inefficiencies in converting disk solids to a protoplanetary core, and subsequently accreting the disk onto the forming planet, would place the disk closer to the Toomre stability limit.  Gl~86~Ab thus presents an important example of either a very massive, albeit truncated, disk, and/or efficient conversion of a protoplanetary disk's mass into a warm Jupiter.

\section{Conclusions} \label{sec:conclusions}

In this study we fit for the orbital parameters of the Gl~86 system based on RV and astrometric data. We find the white dwarf secondary, Gl~86~B, to have a mass \MB % to be \MA, its companion \MB, while 
and the inner planet Gl~86~Ab to have $m\sin{i} = $ \msini. We cannot constrain the inclination and orientation of the inner planet, but obtain good constraints on all other parameters. We obtain an eccentricity of \e and a semimajor axis of \anow for the white dwarf's current orbit. In order to obtain a satisfactory fit to HST astrometry, we require an inflation of $\approx$10\,mas in the relative astrometry uncertainties.  This could also point to an unseen massive companion orbiting Gl~86~B, which would make the system especially unique.  The existence of such a companion might be confirmed or refuted with individual epoch astrometry from a future \gaia data release, since both Gl~86~A and Gl~86~B are detected in \gaia, or by further astrometric monitoring of the system.

Mass loss by the white dwarf Gl~86~B means that its current orbit differs from its primordial orbit.  We combine the MESA initial-final mass relation with a current white dwarf mass of \MB to infer an initial mass of 1.39\,$M_\odot$.  We then run a simulation backward in time to derive the primordial orbit of Gl~86~AB.  We verify that the semimajor axis $a$ satisfies $M_{\rm tot}a =$ constant and eccentricity $e$ satisfies $\left\langle\frac{e}{t}\right\rangle = 0$, in agreement with analytic theory. When both stars were on the main sequence, we infer a semimajor axis of \aprimordial, and an eccentricity matching its current value of \enoerr. This relation assumes isotropic, adiabatic mass loss, as the maximum size of the Gl 86 B was not large enough to initiate Roche lobe overflow.

Finally, we examine how the formation of Gl~86~Ab took place under such a dynamically challenging situation. We find a truncation radius of $\approx$2\,AU for Gl~86~A’s protoplanetary disk, with somewhat smaller values at higher Reynolds numbers, by balancing the viscous stress and inner Lindblad resonant (ILR) torque. We then derive an expression of the total disk mass and infer a total dust mass assuming a dust-to-gas ratio of 1\%. Despite the short separation between the two stars when they were both on the main sequence, the disk mass is sufficient to provide the material to form Gl~86~Ab in the high accretion rate and large Reynolds number (low viscosity) range. This scenario requires an efficient conversion of dust to a planetary core. Inefficient conversion of material would require a more massive disk, which would then approach the Toomre stability limit at its outer truncation radius.

The Gl~86 system, with a $\gtrsim$4\,$\Mjup$ planet formed within a disk truncated at $\approx$2\,AU, demonstrates that giant planet formation is possible even in adverse circumstances.  It shows that severely truncated disks around stars in binaries can birth super-Jovian exoplanets, and provides an important benchmark for planet formation theory.

\acknowledgments
% not yet on a referee comment!
%The authors thank referees for helpful comments. 
Some of this research is based on observations made with the NASA/ESA Hubble Space Telescope obtained from the Space Telescope Science Institute, which is operated by the Association of Universities for Research in Astronomy, Inc., under NASA contract NAS 5–26555. These observations are associated with program 14076. T.D.B.~gratefully  acknowledges  support from the National   Aeronautics and Space Administration (NASA) under grant \#80NSSC18K0439. We acknowledge the traditional owners of the land on which the AAT stands, the Gamilaraay people, and pay our respects to elders past, present, and emerging.

\bibliographystyle{apj_eprint}
\bibliography{refs}

\begin{thebibliography}{}

\bibitem[\protect\citeauthoryear{{Allende Prieto} et~al.}{{Allende Prieto}
  et~al.}{2004}]{Prieto2004}
{Allende Prieto}, C., {Barklem}, P.~S., {Lambert}, D.~L.,  \& {Cunha}, K. 2004,
  \aap, 420, 183

\bibitem[\protect\citeauthoryear{{Artymowicz} \& {Lubow}}{{Artymowicz} \&
  {Lubow}}{1994}]{Artymowicz-Lubow1994}
{Artymowicz}, P.,  \& {Lubow}, S.~H. 1994, \apj, 421, 651

\bibitem[\protect\citeauthoryear{{Barclay} et~al.}{{Barclay}
  et~al.}{2013}]{Barclay2013}
{Barclay}, T., {Rowe}, J.~F., {Lissauer}, J.~J., et~al. 2013, \nat, 494, 452

\bibitem[\protect\citeauthoryear{{Bitsch} et~al.}{{Bitsch}
  et~al.}{2019}]{Bitsch2019}
{Bitsch}, B., {Izidoro}, A., {Johansen}, A., et~al. 2019, \aap, 623, A88

\bibitem[\protect\citeauthoryear{{Bitsch}, {Lambrechts}, \&
  {Johansen}}{{Bitsch} et~al.}{2015}]{Bitsch201510}
{Bitsch}, B., {Lambrechts}, M.,  \& {Johansen}, A. 2015, \aap, 582, A112

\bibitem[\protect\citeauthoryear{{Bitsch} et~al.}{{Bitsch}
  et~al.}{2018}]{Bitsch2018}
{Bitsch}, B., {Morbidelli}, A., {Johansen}, A., et~al. 2018, \aap, 612, A30

\bibitem[\protect\citeauthoryear{{Brandt}}{{Brandt}}{2018}]{Brandt2018}
{Brandt}, T.~D. 2018, \apjs, 239, 31

\bibitem[\protect\citeauthoryear{{Brandt}}{{Brandt}}{2021}]{Brandt2021_HGCA}
{Brandt}, T.~D. 2021, \apjs, 254, 42

\bibitem[\protect\citeauthoryear{{Brandt}, {Dupuy}, \& {Bowler}}{{Brandt}
  et~al.}{2019}]{Brandt2019}
{Brandt}, T.~D., {Dupuy}, T.~J.,  \& {Bowler}, B.~P. 2019, \aj, 158, 140

\bibitem[\protect\citeauthoryear{{Brandt} et~al.}{{Brandt}
  et~al.}{2021}]{Brandt2021}
{Brandt}, T.~D., {Dupuy}, T.~J., {Li}, Y., et~al. 2021, arXiv e-prints,
  arXiv:2105.11671

\bibitem[\protect\citeauthoryear{{Butler} et~al.}{{Butler}
  et~al.}{1997}]{Butler1997}
{Butler}, R.~P., {Marcy}, G.~W., {Williams}, E., {Hauser}, H.,  \& {Shirts}, P.
  1997, \apjl, 474, L115

\bibitem[\protect\citeauthoryear{{Butler} et~al.}{{Butler}
  et~al.}{2006}]{Butler2006}
{Butler}, R.~P., {Wright}, J.~T., {Marcy}, G.~W., et~al. 2006, \apj, 646, 505

\bibitem[\protect\citeauthoryear{{Cantat-Gaudin} \& {Brandt}}{{Cantat-Gaudin}
  \& {Brandt}}{2021}]{Cantat-Gaudin+Brandt_2021}
{Cantat-Gaudin}, T.,  \& {Brandt}, T.~D. 2021, \aap, 649, A124

\bibitem[\protect\citeauthoryear{{Chambers}}{{Chambers}}{2012}]{Chambers2012}
{Chambers}, J.~E. 2012, {Mercury: A software package for orbital dynamics}

\bibitem[\protect\citeauthoryear{{Diego} et~al.}{{Diego}
  et~al.}{1990}]{Diego1990}
{Diego}, F., {Charalambous}, A., {Fish}, A.~C.,  \& {Walker}, D.~D. 1990, in
  Society of Photo-Optical Instrumentation Engineers (SPIE) Conference Series,
  Vol. 1235, Instrumentation in Astronomy VII, ed. D.~L. {Crawford}, 562

\bibitem[\protect\citeauthoryear{{Dosopoulou} \& {Kalogera}}{{Dosopoulou} \&
  {Kalogera}}{2016}]{Dosopoulou2016II}
{Dosopoulou}, F.,  \& {Kalogera}, V. 2016, \apj, 825, 71

\bibitem[\protect\citeauthoryear{{Els} et~al.}{{Els} et~al.}{2001}]{Els2001}
{Els}, S.~G., {Sterzik}, M.~F., {Marchis}, F., et~al. 2001, \aap, 370, L1

\bibitem[\protect\citeauthoryear{{ESA}}{{ESA}}{1997}]{Hipparcos_1997}
{ESA}, ed. 1997, ESA Special Publication, Vol. 1200, {The HIPPARCOS and TYCHO
  catalogues. Astrometric and photometric star catalogues derived from the ESA
  HIPPARCOS Space Astrometry Mission}

\bibitem[\protect\citeauthoryear{{Farihi} et~al.}{{Farihi}
  et~al.}{2013}]{Farihi2013}
{Farihi}, J., {Bond}, H.~E., {Dufour}, P., et~al. 2013, \mnras, 430, 652

\bibitem[\protect\citeauthoryear{{Flock} et~al.}{{Flock}
  et~al.}{2016}]{Flock2016}
{Flock}, M., {Fromang}, S., {Turner}, N.~J.,  \& {Benisty}, M. 2016, \apj, 827,
  144

\bibitem[\protect\citeauthoryear{{Flock} et~al.}{{Flock}
  et~al.}{2019}]{Flock2019}
{Flock}, M., {Turner}, N.~J., {Mulders}, G.~D., et~al. 2019, \aap, 630, A147

\bibitem[\protect\citeauthoryear{{Foreman-Mackey} et~al.}{{Foreman-Mackey}
  et~al.}{2013}]{Foreman-Mackey_2013}
{Foreman-Mackey}, D., {Hogg}, D.~W., {Lang}, D.,  \& {Goodman}, J. 2013, \pasp,
  125, 306

\bibitem[\protect\citeauthoryear{{Fuhrmann} et~al.}{{Fuhrmann}
  et~al.}{2014}]{Fuhrmann2014}
{Fuhrmann}, K., {Chini}, R., {Buda}, L.~S.,  \& {Pozo Nu{\~n}ez}, F. 2014,
  \apj, 785, 68

\bibitem[\protect\citeauthoryear{{Gaia Collaboration} et~al.}{{Gaia
  Collaboration} et~al.}{2021}]{Gaia-EDR3-summary}
{Gaia Collaboration}, {Brown}, A.~G.~A., {Vallenari}, A., et~al. 2021, \aap,
  649, A1

\bibitem[\protect\citeauthoryear{{Geller} et~al.}{{Geller}
  et~al.}{2021}]{Geller..Mathieu..2021}
{Geller}, A.~M., {Mathieu}, R.~D., {Latham}, D.~W., et~al. 2021, arXiv
  e-prints, arXiv:2101.07883

\bibitem[\protect\citeauthoryear{{Gong} \& {Ji}}{{Gong} \&
  {Ji}}{2018}]{Gong2018}
{Gong}, Y.-X.,  \& {Ji}, J. 2018, \mnras, 478, 4565

\bibitem[\protect\citeauthoryear{{Grevesse} et~al.}{{Grevesse}
  et~al.}{2012}]{Grevesse2012}
{Grevesse}, N., {Asplund}, M., {Sauval}, A.~J.,  \& {Scott}, P. 2012, in
  Astronomical Society of the Pacific Conference Series, Vol. 462, Progress in
  Solar/Stellar Physics with Helio- and Asteroseismology, ed. H.~{Shibahashi},
  M.~{Takata}, \& A.~E. {Lynas-Gray}, 41

\bibitem[\protect\citeauthoryear{{Guillot}, {Ida}, \& {Ormel}}{{Guillot}
  et~al.}{2014}]{Guillot2014}
{Guillot}, T., {Ida}, S.,  \& {Ormel}, C.~W. 2014, \aap, 572, A72

\bibitem[\protect\citeauthoryear{{Hadjidemetriou}}{{Hadjidemetriou}}{1963}]{Hadjidemetriou1963}
{Hadjidemetriou}, J.~D. 1963, \icarus, 2, 440

\bibitem[\protect\citeauthoryear{{Halbwachs} et~al.}{{Halbwachs}
  et~al.}{2000}]{Halbwachs+Arenou+Mayor+etal_2000}
{Halbwachs}, J.~L., {Arenou}, F., {Mayor}, M., {Udry}, S.,  \& {Queloz}, D.
  2000, \aap, 355, 581

\bibitem[\protect\citeauthoryear{{Hatzes} et~al.}{{Hatzes}
  et~al.}{2003}]{Hatzes2003}
{Hatzes}, A.~P., {Cochran}, W.~D., {Endl}, M., et~al. 2003, \apj, 599, 1383

\bibitem[\protect\citeauthoryear{{Hayashi}}{{Hayashi}}{1981}]{Hayashi_1981}
{Hayashi}, C. 1981, Progress of Theoretical Physics Supplement, 70, 35

\bibitem[\protect\citeauthoryear{{Henry} et~al.}{{Henry}
  et~al.}{2000}]{Henry2000}
{Henry}, G.~W., {Marcy}, G.~W., {Butler}, R.~P.,  \& {Vogt}, S.~S. 2000, \apjl,
  529, L41

\bibitem[\protect\citeauthoryear{{Jang-Condell}}{{Jang-Condell}}{2015}]{Jang-Condell2015}
{Jang-Condell}, H. 2015, \apj, 799, 147

\bibitem[\protect\citeauthoryear{{Jang-Condell}, {Mugrauer}, \&
  {Schmidt}}{{Jang-Condell} et~al.}{2008}]{Jang-Condell2008}
{Jang-Condell}, H., {Mugrauer}, M.,  \& {Schmidt}, T. 2008, \apjl, 683, L191

\bibitem[\protect\citeauthoryear{{Jang-Condell} \& {Sasselov}}{{Jang-Condell}
  \& {Sasselov}}{2003}]{Jang-Condell2003}
{Jang-Condell}, H.,  \& {Sasselov}, D.~D. 2003, \apj, 593, 1116

\bibitem[\protect\citeauthoryear{{Jeans}}{{Jeans}}{1924}]{Jeans1924}
{Jeans}, J.~H. 1924, \mnras, 85, 2

\bibitem[\protect\citeauthoryear{{Jenkins} et~al.}{{Jenkins}
  et~al.}{2015}]{Jenkins2015}
{Jenkins}, J.~M., {Twicken}, J.~D., {Batalha}, N.~M., et~al. 2015, \aj, 150, 56

\bibitem[\protect\citeauthoryear{{Johansen} \& {Lambrechts}}{{Johansen} \&
  {Lambrechts}}{2017}]{Johansen-Lambrechts2017}
{Johansen}, A.,  \& {Lambrechts}, M. 2017, Annual Review of Earth and Planetary
  Sciences, 45, 359

\bibitem[\protect\citeauthoryear{{Jones} et~al.}{{Jones}
  et~al.}{2010}]{Jones2010}
{Jones}, H. R.~A., {Butler}, R.~P., {Tinney}, C.~G., et~al. 2010, \mnras, 403,
  1703

\bibitem[\protect\citeauthoryear{{Kalirai} et~al.}{{Kalirai}
  et~al.}{2008}]{Kalirai2008}
{Kalirai}, J.~S., {Hansen}, B. M.~S., {Kelson}, D.~D., et~al. 2008, \apj, 676,
  594

\bibitem[\protect\citeauthoryear{{Kama}, {Min}, \& {Dominik}}{{Kama}
  et~al.}{2009}]{Kama2009}
{Kama}, M., {Min}, M.,  \& {Dominik}, C. 2009, \aap, 506, 1199

\bibitem[\protect\citeauthoryear{{Kostov} et~al.}{{Kostov}
  et~al.}{2021}]{Kostov2021}
{Kostov}, V.~B., {Powell}, B.~P., {Orosz}, J.~A., et~al. 2021, \aj, 162, 234

\bibitem[\protect\citeauthoryear{{Kratter} \& {Perets}}{{Kratter} \&
  {Perets}}{2012}]{Kratter-Perets2012}
{Kratter}, K.~M.,  \& {Perets}, H.~B. 2012, \apj, 753, 91

\bibitem[\protect\citeauthoryear{{Kraus} et~al.}{{Kraus}
  et~al.}{2016}]{Kraus2016}
{Kraus}, A.~L., {Ireland}, M.~J., {Huber}, D., {Mann}, A.~W.,  \& {Dupuy},
  T.~J. 2016, \aj, 152, 8

\bibitem[\protect\citeauthoryear{{Lagrange} et~al.}{{Lagrange}
  et~al.}{2006}]{Lagrange2006}
{Lagrange}, A.~M., {Beust}, H., {Udry}, S., {Chauvin}, G.,  \& {Mayor}, M.
  2006, arXiv e-prints, astro

\bibitem[\protect\citeauthoryear{{Lam} et~al.}{{Lam} et~al.}{2020}]{Lam2020}
{Lam}, K. W.~F., {Korth}, J., {Masuda}, K., et~al. 2020, \aj, 159, 120

\bibitem[\protect\citeauthoryear{{Lambrechts} \& {Johansen}}{{Lambrechts} \&
  {Johansen}}{2012}]{Lambrechts-Johansen2012}
{Lambrechts}, M.,  \& {Johansen}, A. 2012, \aap, 544, A32

\bibitem[\protect\citeauthoryear{{Leahy} \& {Leahy}}{{Leahy} \&
  {Leahy}}{2015}]{Leahy2015}
{Leahy}, D.~A.,  \& {Leahy}, J.~C. 2015, Computational Astrophysics and
  Cosmology, 2, 4

\bibitem[\protect\citeauthoryear{{Lindegren} et~al.}{{Lindegren}
  et~al.}{2021}]{GaiaEDR3-ast-sol}
{Lindegren}, L., {Klioner}, S.~A., {Hern{\'a}ndez}, J., et~al. 2021, \aap, 649,
  A2

\bibitem[\protect\citeauthoryear{{Luger} et~al.}{{Luger}
  et~al.}{2017}]{Luger2017}
{Luger}, R., {Sestovic}, M., {Kruse}, E., et~al. 2017, Nature Astronomy, 1,
  0129

\bibitem[\protect\citeauthoryear{{Marcy} \& {Butler}}{{Marcy} \&
  {Butler}}{2000}]{Marcy+Butler_2000}
{Marcy}, G.~W.,  \& {Butler}, R.~P. 2000, \pasp, 112, 137

\bibitem[\protect\citeauthoryear{{Meibom} \& {Mathieu}}{{Meibom} \&
  {Mathieu}}{2005}]{Meibom..Mathieu..2005}
{Meibom}, S.,  \& {Mathieu}, R.~D. 2005, \apj, 620, 970

\bibitem[\protect\citeauthoryear{{Mugrauer} \& {Neuh{\"a}user}}{{Mugrauer} \&
  {Neuh{\"a}user}}{2005}]{Mugrauer2005}
{Mugrauer}, M.,  \& {Neuh{\"a}user}, R. 2005, \mnras, 361, L15

\bibitem[\protect\citeauthoryear{{Natta} et~al.}{{Natta}
  et~al.}{2001}]{Natta2001}
{Natta}, A., {Prusti}, T., {Neri}, R., et~al. 2001, \aap, 371, 186

\bibitem[\protect\citeauthoryear{{Orosz} et~al.}{{Orosz}
  et~al.}{2019}]{Orosz2019}
{Orosz}, J.~A., {Welsh}, W.~F., {Haghighipour}, N., et~al. 2019, \aj, 157, 174

\bibitem[\protect\citeauthoryear{{Paardekooper} \& {Leinhardt}}{{Paardekooper}
  \& {Leinhardt}}{2010}]{Paardekooper2010}
{Paardekooper}, S.~J.,  \& {Leinhardt}, Z.~M. 2010, \mnras, 403, L64

\bibitem[\protect\citeauthoryear{{Paxton} et~al.}{{Paxton}
  et~al.}{2011}]{Paxton2011}
{Paxton}, B., {Bildsten}, L., {Dotter}, A., et~al. 2011, \apjs, 192, 3

\bibitem[\protect\citeauthoryear{{Pollack} et~al.}{{Pollack}
  et~al.}{1996}]{Pollack1996}
{Pollack}, J.~B., {Hubickyj}, O., {Bodenheimer}, P., et~al. 1996, \icarus, 124,
  62

\bibitem[\protect\citeauthoryear{{Queloz} et~al.}{{Queloz}
  et~al.}{2000}]{Queloz2000}
{Queloz}, D., {Mayor}, M., {Weber}, L., et~al. 2000, \aap, 354, 99

\bibitem[\protect\citeauthoryear{{Rafikov} \& {Silsbee}}{{Rafikov} \&
  {Silsbee}}{2015}]{Rafikov2015}
{Rafikov}, R.~R.,  \& {Silsbee}, K. 2015, \apj, 798, 70

\bibitem[\protect\citeauthoryear{{Ram{\'\i}rez}, {Allende Prieto}, \&
  {Lambert}}{{Ram{\'\i}rez} et~al.}{2007}]{Ramirez2007}
{Ram{\'\i}rez}, I., {Allende Prieto}, C.,  \& {Lambert}, D.~L. 2007, \aap, 465,
  271

\bibitem[\protect\citeauthoryear{{Ram{\'\i}rez}, {Allende Prieto}, \&
  {Lambert}}{{Ram{\'\i}rez} et~al.}{2013}]{Ramirez2013}
{Ram{\'\i}rez}, I., {Allende Prieto}, C.,  \& {Lambert}, D.~L. 2013, \apj, 764,
  78

\bibitem[\protect\citeauthoryear{{Ramm} et~al.}{{Ramm} et~al.}{2016}]{Ramm2016}
{Ramm}, D.~J., {Nelson}, B.~E., {Endl}, M., et~al. 2016, \mnras, 460, 3706

\bibitem[\protect\citeauthoryear{{Safronov}}{{Safronov}}{1960}]{Safronov_1960}
{Safronov}, V.~S. 1960, Annales d'Astrophysique, 23, 979

\bibitem[\protect\citeauthoryear{{Santos}, {Israelian}, \& {Mayor}}{{Santos}
  et~al.}{2000}]{Santos2000}
{Santos}, N.~C., {Israelian}, G.,  \& {Mayor}, M. 2000, \aap, 363, 228

\bibitem[\protect\citeauthoryear{{Santos}, {Israelian}, \& {Mayor}}{{Santos}
  et~al.}{2004}]{Santos2004}
{Santos}, N.~C., {Israelian}, G.,  \& {Mayor}, M. 2004, \aap, 415, 1153

\bibitem[\protect\citeauthoryear{{Shakura} \& {Sunyaev}}{{Shakura} \&
  {Sunyaev}}{1973}]{Shakura-Sunyaev1973}
{Shakura}, N.~I.,  \& {Sunyaev}, R.~A. 1973, \aap, 500, 33

\bibitem[\protect\citeauthoryear{{Shallue} \& {Vanderburg}}{{Shallue} \&
  {Vanderburg}}{2018}]{Shallue2018}
{Shallue}, C.~J.,  \& {Vanderburg}, A. 2018, \aj, 155, 94

\bibitem[\protect\citeauthoryear{{Smith} et~al.}{{Smith}
  et~al.}{2018}]{Smith2018}
{Smith}, A.~M.~S., {Cabrera}, J., {Csizmadia}, S., et~al. 2018, \mnras, 474,
  5523

\bibitem[\protect\citeauthoryear{{Su} et~al.}{{Su} et~al.}{2021}]{Su2021}
{Su}, X.-N., {Xie}, J.-W., {Zhou}, J.-L.,  \& {Thebault}, P. 2021, arXiv
  e-prints, arXiv:2109.14577

\bibitem[\protect\citeauthoryear{{Teske} et~al.}{{Teske}
  et~al.}{2016}]{Teske2016}
{Teske}, J.~K., {Shectman}, S.~A., {Vogt}, S.~S., et~al. 2016, \aj, 152, 167

\bibitem[\protect\citeauthoryear{{Tinney} et~al.}{{Tinney}
  et~al.}{2001}]{Tinney2001}
{Tinney}, C.~G., {Butler}, R.~P., {Marcy}, G.~W., et~al. 2001, \apj, 551, 507

\bibitem[\protect\citeauthoryear{{Toomre}}{{Toomre}}{1964}]{Toomre_1964}
{Toomre}, A. 1964, \apj, 139, 1217

\bibitem[\protect\citeauthoryear{{Ueda}, {Okuzumi}, \& {Flock}}{{Ueda}
  et~al.}{2017}]{Ueda2017}
{Ueda}, T., {Okuzumi}, S.,  \& {Flock}, M. 2017, \apj, 843, 49

\bibitem[\protect\citeauthoryear{{van Leeuwen}}{{van
  Leeuwen}}{2007}]{vanLeeuwen_2007}
{van Leeuwen}, F. 2007, \aap, 474, 653

\bibitem[\protect\citeauthoryear{{Venner}, {Vanderburg}, \& {Pearce}}{{Venner}
  et~al.}{2021}]{Venner2021}
{Venner}, A., {Vanderburg}, A.,  \& {Pearce}, L.~A. 2021, \aj, 162, 12

\bibitem[\protect\citeauthoryear{{Veras}, {Hadjidemetriou}, \& {Tout}}{{Veras}
  et~al.}{2013}]{Veras2013}
{Veras}, D., {Hadjidemetriou}, J.~D.,  \& {Tout}, C.~A. 2013, \mnras, 435, 2416

\bibitem[\protect\citeauthoryear{{Vousden}, {Farr}, \& {Mandel}}{{Vousden}
  et~al.}{2016}]{Vousden+Farr+Mandel_2016}
{Vousden}, W.~D., {Farr}, W.~M.,  \& {Mandel}, I. 2016, \mnras, 455, 1919

\bibitem[\protect\citeauthoryear{{Wang} et~al.}{{Wang} et~al.}{2014}]{Wang2014}
{Wang}, J., {Xie}, J.-W., {Barclay}, T.,  \& {Fischer}, D.~A. 2014, \apj, 783,
  4

\bibitem[\protect\citeauthoryear{{Welsh} et~al.}{{Welsh}
  et~al.}{2012}]{Welsh2012}
{Welsh}, W.~F., {Orosz}, J.~A., {Carter}, J.~A., et~al. 2012, \nat, 481, 475

\bibitem[\protect\citeauthoryear{{Wittenmyer} et~al.}{{Wittenmyer}
  et~al.}{2014}]{Wittenmyer2014}
{Wittenmyer}, R.~A., {Horner}, J., {Tinney}, C.~G., et~al. 2014, \apj, 783, 103

\end{thebibliography}
\clearpage

\end{document}